%% file: ms.tex
\documentclass[preprint2]{aastex}
\usepackage{epsfig}
\usepackage{natbib}

\begin{document}
\input{defs}
\title{Our Peculiar Motion Away from the Local Void}

\author{R. Brent Tully,}
\affil{Institute for Astronomy, University of Hawaii, 2680 Woodlawn Drive,
 Honolulu, HI 96822}

\and

\author{Edward J. Shaya}
\affil{University of Maryland, Astronomy Department, College Park, MD 20743}

\and

\author{Igor D. Karachentsev.}
\affil{Special Astrophysical Observatory, Nizhnij Arkhyz, Karachaevo-Cherkessia, Russia}

\and

\author{H\'el\`ene M. Courtois, Dale D. Kocevski, and Luca Rizzi}
\affil{Institute for Astronomy, University of Hawaii, 2680 Woodlawn Drive,
 Honolulu, HI 96822}

\and

\author{Alan Peel}
\affil{University of Maryland, Astronomy Department, College Park, MD 20743}

\begin{abstract}
The peculiar velocity of the Local Group of galaxies manifested in the Cosmic Microwave Background dipole is found to decompose into three dominant components.  The three components are clearly separated because they arise on distinct spatial scales and are fortuitously almost orthogonal in their influences.  The nearest, which is distinguished by a velocity discontinuity at $\sim 7$ Mpc, arises from the evacuation of the Local Void.  We lie in the Local Sheet that bounds the void.  Random motions within the Local Sheet are small and we advocate a reference frame with respect to the Local Sheet in preference to the Local Group.  Our Galaxy participates in the bulk motion of the Local Sheet away from the Local Void.  The component of our motion on an intermediate scale is attributed to the Virgo Cluster and its surroundings, 17 Mpc away.  The third and largest component is an attraction on scales larger than 3000~\kms\ and centered near the direction of the Centaurus Cluster.  The amplitudes of the three components are 259, 185, and 455 \kms, respectively, adding collectively to 631~\kms\ in the reference frame of the Local Sheet.    Taking the nearby influences into account, particularly that of the Local Void, causes the residual attributed to large scales to align with observed concentrations of distant galaxies and reduces somewhat the amplitude of motion attributed to their pull.    Turning to small scales, in addition to the motion of our Local Sheet away from the Local Void, the nearest adjacent filament, the Leo Spur, is seen to be moving in a direction that will lead to convergence with our filament. Finally, a good distance to an isolated galaxy within the Local Void reveals that this dwarf system has a substantial motion of at least 230~\kms\  away from the void center.  Given the velocities expected from gravitational instability theory in the standard cosmological paradigm, the distance to the center of the Local Void  must be at least 23 Mpc from our position, implying the Local Void is extremely large.

\end{abstract}

\keywords{galaxies: distances and redshifts -- dark matter -- large scale structure of the universe}

\section{Introduction}

The dipole anisotropy seen in the cosmic microwave background (CMB) temperature map 
\citep{1996ApJ...473..576F} is compelling evidence that the solar system has a large peculiar motion with respect to the overall cosmic expansion.  There are known local components to this motion, including the orbital velocity of the Sun in the Milky Way Galaxy and the attraction of our Galaxy toward M31.  Once these components are taken into account, it is found that the Local Group of galaxies has a peculiar motion of over 600 \kms\  in a well established direction. 

Soon after the discovery of the CMB dipole the coincidence  in direction of our motion with prominent large scale structure was noted \citep{1984ApJ...280..470S} and then evidence was found for flows of nearby galaxies toward this direction \citep{1988ApJ...326...19L}.  There has been great interest in trying to identify the dominant source, or at least the characteristic distance, of the `great attractor' causing our large scale motion.    This interest is well summarized in several conference proceedings 
\citep{1988lsmu.book.....R, 1993cvf..conf.....B, 2000ASPC..201.....C}.
The issue has been complicated by the observation that two important structures lie in the general direction of our motion: the Norma-Hydra-Centaurus complex in the foreground and the
enormous Shapley Concentration in the background  \citep{1989Natur.338..562S, 1989Natur.342..251R}.  The debate continues regarding the relative importance of these structures on our motion
\citep{2006ApJ...645.1043K, 2006MNRAS.368.1515E}.

There has been a long-standing appreciation that there are significant dynamical influences on intermediate scales within what has traditionally been called the Local Supercluster.  Our galaxy is known to experience a pull toward the Virgo Cluster at the heart of the Local Supercluster
 \citep{1981ApJ...246..680T, 1982ApJ...258...64A, 1982ApJ...263..485H, 1984ApJ...281...31T, 2000ApJ...530..625T}.  However, the story on intermediate scales is more complicated than just an attraction centered on or near the Virgo Cluster.  The Numerical Action Method (NAM) models of \citet{1995ApJ...454...15S} assign mass according to the complex distribution of light and provide a reasonable description of galaxy motions.  Still, NAM reconstructions have not yet provided a fully satisfactory explanation of the `local velocity anomaly'  \citep{1988lsmu.book..169T, 1988lsmu.book..115F, 
1992ApJS...80..479T}.  We use the term to describe the pattern of negative motions with respect to Hubble expansion of galaxies in a neighboring filament called the Leo Spur in the Nearby Galaxies Atlas \citep{1987nga..book.....T}.

In the present paper we return to the problem of the local velocity anomaly.   Imaging with Hubble Space Telescope (HST) has provided a wealth of accurate distances to nearby galaxies based on measures of the luminosity of stars at the tip of the red giant branch, the TRGB method   \citep{2004AJ....127.2031K, 2006AJ....131.1361K}.  Additionally, over the years many other good distances have become available.  Those that are particularly important for this work include those provided by the HST Cepheid Key Project \citep{2001ApJ...553...47F}, the Surface Brightness Fluctuation (SBF) study of 
\citet{2001ApJ...546..681T}, and two catalogs of luminosity--linewidth distances, one a sample of extreme edge-on galaxies with 2MASS magnitudes that has been discussed by  \citet{2002A&A...396..431K} and the other an extension of the sample discussed by \citet{2000ApJ...533..744T}.  These new observations of distances have clarified that the phenomenon referred to as the `local velocity anomaly' definitely exists but it is so much more extensive than previously suspected that the adjective `local' may not be appropriate.  It will be shown that the observed anomalous motion has nothing to do with the known pull toward the Virgo Cluster nor to the large--scale great attractor(s).

\section{A Catalog of Galaxy Distances}

\begin{figure}
\figurenum{1}
\centering
\includegraphics[scale=0.3]{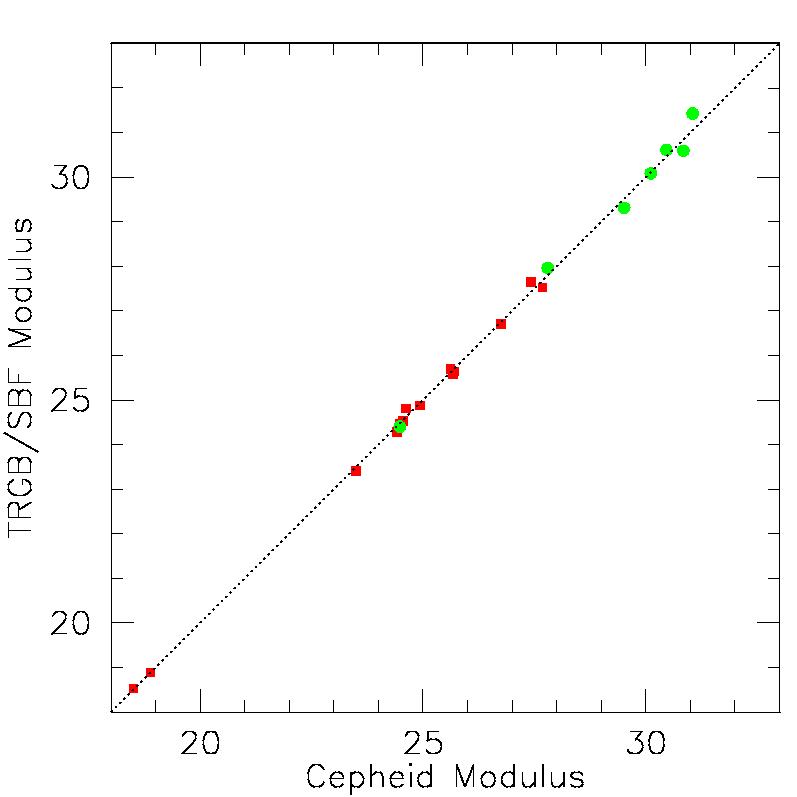}
\includegraphics[scale=0.3]{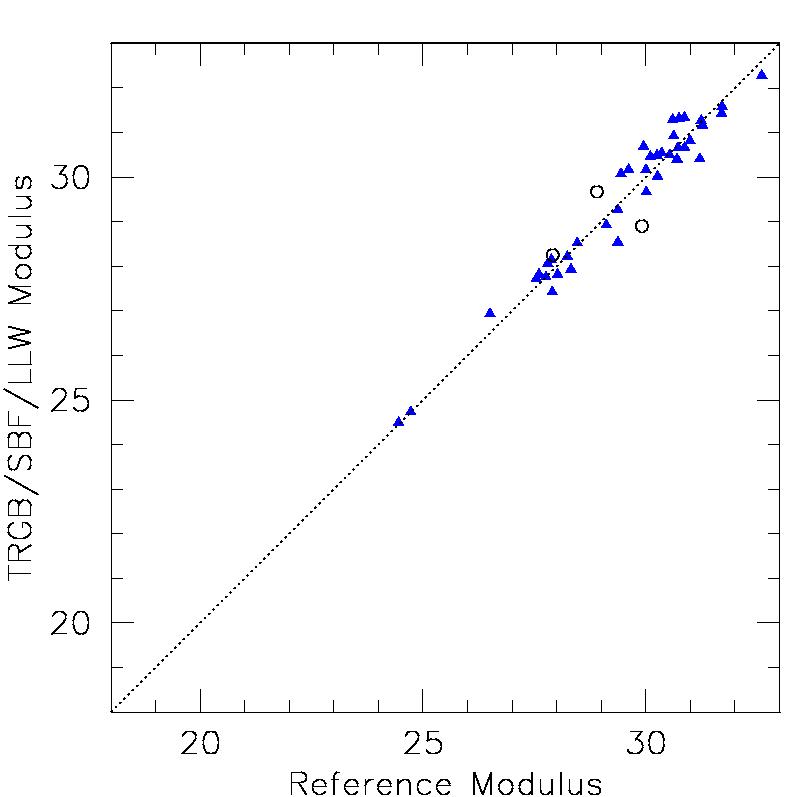}
\includegraphics[scale=0.3]{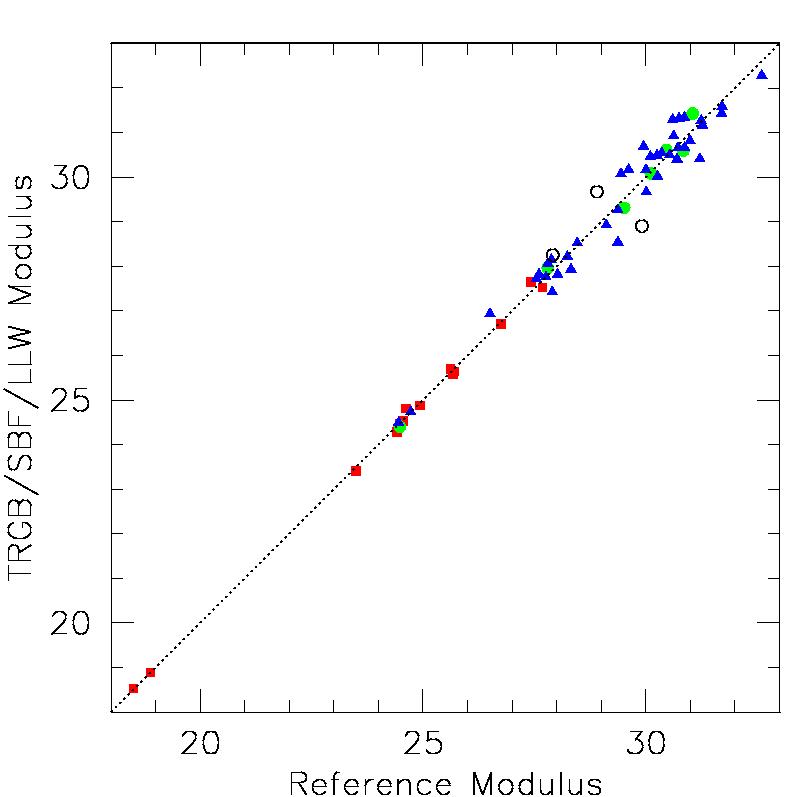}
\caption{Comparison of distance moduli determined by different methods.  {\bf Top:}  Comparison of TRGB moduli (red squares) and SBF moduli (green circles) with moduli determined with Cepheid variables.  {\bf Middle:} Comparison of luminosity--linewidth moduli with moduli determined by either Cepheids or the TRGB.  Source: Tully (blue triangles) and Karachentsev (black open circles)  {\bf Bottom:} Combination of the top two panels. }
\label{compare}
\end{figure}

Our database is an outgrowth of the Nearby Galaxies Catalog \citep{1988ngc..book.....T} and for the
current discussion has the same limit of 3,000 \kms.  We presently have distance estimates for 1791 galaxies in 743 groups in this volume derived from four different methods. 
The reference scale for our distances is set by the HST Cepheid Key Project observations  \citep{2001ApJ...553...47F}.  Including all sources, we have 51 distances by the Cepheid method.  Next we add galaxies with TRGB distance estimates.  Individual TRGB distances are of comparable quality to the Cepheid values and are demonstrated to be on a consistent scale \citep{2003AJ....125.1261D,
2004ApJ...608...42S, 2007ApJ...661..815R}.  Procedures for measuring the TRGB are discussed by \cite{1996ApJ...461..713S} and \citet{2006AJ....132.2729M}.  Distance moduli are directly compared in the top panel of Figure~\ref{compare} for 14 galaxies with both Cepheid and TRGB measurements.  There are 221 TRGB estimates in the present sample.  Third, we accept the SBF measures of \citet{2001ApJ...546..681T} and \citet{2007ApJ...655..144M}.  The Tonry measures are available for 299 galaxies around the sky while the Mei sample of 84 galaxies is restricted to the Virgo Cluster and a projected group.  The claimed accuracies with SBF are comparable with the Cepheid and TRGB accuracies.  The zero point for SBF distances is confirmed to agree with the Cepheid scale.  The top panel of Figure~\ref{compare} compares distance moduli for 7 galaxies with both Cepheid and SBF determinations.  Of course, the TRGB and SBF methods are intimately related; both use the standard candle nature of the brightest red giant stars.  The TRGB method requires that the stars be individually resolved and hence can only be applied to nearby galaxies.  The SBF method uses the statistical properties of these stars as they blend together in an image and can be applied to larger distances.  Some galaxies have been observed by more than one of the above three methods.  In total, 601 galaxies in our sample have been observed by at least one of the Cepheid, TRGB, or SBF methods.  We assign an uncertainty of 10\% to a distance obtained by one of these methods.  Generally, the galaxies with TRGB estimates lie within 7 Mpc and the galaxies with SBF estimates lie beyond this distance.

%
%

On top of these, we add two luminosity--linewidth samples.  The larger of these involves 1030 distance measures derived from the correlation between galaxy luminosity and rotation rate as measured from the width of an HI line profile \citep{1977A&A....54..661T}.  The calibration is that of \citet{2000ApJ...533..744T} shifted slightly to be consistent in zero point with the HST  Cepheid Key Project results.  This zero point is set by 40 galaxies with Cepheid or TRGB distance measures.  Here is our current calibration:
\begin{equation}
M_B^{b,i,k} = -19.99 -7.27 (W_R^i - 2.5)
\end{equation}
\begin{equation}
M_R^{b,i,k} = -21.00 -7.65 (W_R^i - 2.5)
\end{equation}
\begin{equation}
M_I^{b,i,k} = -21.43 -8.11 (W_R^i - 2.5)
\end{equation}
\begin{equation}
M_H^{b,i,k} = -22.17 -9.55 (W_R^i - 2.5)
\end{equation}
where the superscripts on the $B,R,I,H$ absolute magnitudes indicate corrections have been made for obscuration within our Galaxy ($b$) and due to the inclination of the target galaxy ($i$) and for redshift effects ($k$)   \citep{2000ApJ...533..744T}.  The parameter $W_R^i$ is a measure of the inclination corrected neutral Hydrogen linewidth  \citep{1985ApJS...58...67T}.  The optical band magnitudes $B,R,I$ are `total' values.  The near infrared $H$ magnitudes are aperture values in the system of  \citet{1986ApJ...302..536A}.  

The other luminosity--linewidth sample is composed of edge-on galaxies with 2MASS $K$-band photometry \citep{2002A&A...396..431K} restricted to less than 3,000~\kms.  This sample contributes 402 distances, 178 already included.  The substantial overlap between the two luminosity--linewidth samples provides
confirmation that the zero points are the same and gives rms agreement per measure of 0.39 mag.  The excellent agreement in distance moduli between the luminosity--linewidth and other measures is shown in the middle panel of Figure~\ref{compare}.  The bottom panel of Figure~\ref{compare} compares all cases in the current sample with distance measurements by more than one method. 

The luminosity--linewidth distance estimates are considered to have an accuracy of 20\% rms for a single observation.  They are less accurate than those obtained with the procedures previously discussed but are much more numerous.  SBF observations are restricted to early--type galaxies that tend to reside together in high density environments.  Luminosity--linewidth observations are restricted to spiral galaxies that are more widely distributed.  The combination of the two provides a rich sampling of the distribution of galaxies and their motions throughout the Local Supercluster.

Our current database of distances for galaxies within 3,000~\kms\ is provided in two tables that can be accessed in their entirety in electronic form.  Table~1 identifies the 1791 individual galaxies with measured distances.  The column information is as follows. (1) J2000 equatorial coordinates.
(2) Principal Galaxies Catalogue (PGC) name from the Lyon Extragalactic Database (LEDA:
http://leda.univ-lyon1.fr/). (3) Common name. (4) Group ID for cross-reference with Table 2.  (5) NBG ID, the group ID in the Nearby Galaxies Catalog  \citep{1988ngc..book.....T}. (6,7) Galactic longitude and latitude. (8,9) Supergalactic longitude and latitude. (10) Numeric morphological type code. (11) Differential Galactic reddening $E(B-V)$ \citep{1998ApJ...500..525S}.  (12) Total blue magnitudes, mostly from the Third Reference Catalogue \citep{1991trcb.book.....D}. (13,14,15,16) Velocities in the reference frames of the Sun, the Galactic center, the Local Sheet (defined later), and the CMB, in \kms. The columns 17 to 29 are filled if the galaxy has a luminosity--linewidth distance estimate based on the revised \citet{2000ApJ...533..744T} calibration. (17) Photometrically derived ratio of minor to major axes $b/a$, related to the galaxy inclination $i$ by ${\rm cos}~i = [((b/a)^2 - q_0^2))/(1-q_0^2)]^{1/2}$ where $q_0=0.2$ is taken as the axial ratio of a spiral galaxy seen edge on. (18) Number of sources for the measurement of axial ratio. (19) Total $B$ magnitude from CCD area photometry.  (20,21) Total $R$ magnitude and number of sources for $R$ magnitude.  (22,23) Total $I$ magnitude and number of sources for $I$ magnitude. (24,25) $H_{-0.5}$ aperture magnitude and number of sources of $H$ photometry.  (26,27) Heliocentric velocity and linewidth based on HI observations; the linewidth is the parameter $W_R^i$ defined by \citet{1985ApJS...58...67T}, including rectification from the viewing inclination to edge on, defined to agree statistically with twice the maximum rotation velocity.  (28,29) Distance modulus and uncertainty determined from the luminosity--linewidth method, where the uncertainty reflects a weighting of the separate bandpasses.  (30) Distance modulus determined in the case of a galaxy from the flat galaxy--2MASS sample;  an uncertainty of
0.40 mag is accepted in these cases.  (31,32) Distance modulus given by either the Surface Brightness Fluctuation method ($s$), the brightest Red Giant Branch stars ($r$), or the Cepheid period--luminosity relation ($c$), and indication of the source, $s$, $r$, or $c$; an uncertainty of 0.2 mag is accepted in these cases.  Table~1 is available at 

\noindent
ifa.hawaii.edu/$\sim$tully/voidtable1.

In Table~2 information is reassembled and averaged within 743 groups (including groups  of one).  The columns are described below.  (1) A unique group identification number; appears in column 4 of Table~1 for individual galaxies.  (2) NBG ID, as in column 5 of Table~1.  (3,4) Galactic longitude and latitude of group.  (5,6) Supergalactic longitude and latitude of group.  (7) Logarithm of B absolute luminosity summed over group and based on observed distance. (8,9,10,11) Group averaged velocities in
reference frames of the Sun, the Galactic center, the Local Sheet, and the CMB, in \kms.  (12,13) Distance modulus and uncertainty, averaged over all estimates for group members.  (14,15,16,17) Distance, and components of distance in the Supergalactic SGX, SGY, and SGZ directions, in Mpc.  (18) Peculiar velocity if H$_0=74$~\kmsMpc, $V_{pec} = V_{LS} - {\rm H}_0 d$, in \kms, where $V_{LS}$ is the velocity in column 10 and $d$ is the distance in column 14.  (19,20,21) Number of galaxies in group with luminosity--linewidth distance measures from the extended Tully--Pierce sample, the averaged distance modulus from luminosity--linewidth measures, and the assigned uncertainty.  (22,23,24) Number of galaxies in group with distances measures from the flat galaxies--2MASS sample, the averaged modulus, and uncertainty.  (25,26,27) The sum of the number of galaxies in the group with Surface Brightness Fluctuation, Tip of the Red Giant Branch, or Cepheid distance measures, the averaged modulus, and uncertainty. 
Table~2 is available at 

\noindent
ifa.hawaii.edu/$\sim$tully/voidtable2-743groups.

\clearpage
\begin{deluxetable}{lrlrlrrrrrrrrrrrrrrrrrrrrrrrrrrc}
\tabletypesize{\scriptsize}
\rotate
\tablewidth{0pt}
\tablenum{1}
\tablecolumns{32}
\tablecaption{Distance Estimates for 1791 Galaxies}
\label{tbl:dist_indiv}

\tablehead{\colhead{1}&\colhead{2}&\colhead{3}&\colhead{4}&\colhead{5}&\colhead{6}&\colhead{7}&\colhead{8}&\colhead{9}&\colhead{10}&\colhead{11}&\colhead{12}&\colhead{13}&\colhead{14}&\colhead{15}&\colhead{16}&\colhead{17}&\colhead{18}&\colhead{19}&\colhead{20}&\colhead{21}&\colhead{22}&\colhead{23}&\colhead{24}&\colhead{25}&\colhead{26}&\colhead{27}&\colhead{28}&\colhead{29}&\colhead{30}&\colhead{31}&\colhead{32}}
\tablehead{\colhead{RA (J2000) Dec}&\colhead{PGC}&\colhead{Name}&\colhead{Gp ID}&\colhead{NBG ID}&\colhead{$\ell$}&\colhead{$b$}&\colhead{SGL}&\colhead{SGB}&\colhead{T}&\colhead{$E(B-V)$}&\colhead{$B_T$}&\colhead{$V_{\odot}$}&\colhead{$V_{GSR}$}&\colhead{$V_{LS}$}&\colhead{$V_{CMB}$}&\colhead{$b/a$}&\colhead{N$_i$}&\colhead{$B$}&\colhead{$R$}&\colhead{N$_R$}&\colhead{$I$}&\colhead{N$_I$}&\colhead{$H$}&\colhead{N$_H$}&\colhead{$V_{21}$}&\colhead{$W_R^i$}&\colhead{$\mu_{LL}$}&\colhead{$\epsilon_{LL}$}&\colhead{$\mu_{FG}$}&\colhead{$\mu_{src}$}&\colhead{$src$}}



\startdata
000158.5-152741 &    143& WLM       &   222& 14-12 12&   75.8655& -73.6256& 277.8076&   8.0847&  10& 0.036& 11.04&  -127&  -74&  -33& -462&       &  &      &      &  &      &  &      &  &     &    &        &     &        & 24.89& ~rc\\
000315.0+160843 &    218& NGC7814   &  1211& 65 ~-6  ~6&  106.4094& -45.1749& 309.0612&  16.4021&   2& 0.045& 11.57&  1054& 1194& 1279&  696&   0.19& 1& 11.72& 10.01& 1&  9.40& 1&  0.00& 0& 1054& 521&   31.29& 0.36&        & 30.60& s~~\\
000358.7+204506 &    279& NGC7817   &  1178& 64 ~-8  ~8&  108.2271& -40.7610& 313.8132&  17.1426&   4& 0.058& 12.74&  2308& 2457& 2547& 1956&   0.27& 3&  0.00&  0.00& 0& 10.60& 3&  9.25& 1& 2308& 432&   32.10& 0.38&  32.01 &      &    \\
000620.1-412945 &    474& ESO293-034&  1088& 61  ~~0 16&  332.8271& -72.9123& 253.5419&  -1.5693&   6& 0.017& 13.64&  1516& 1482& 1474& 1278&       &  &      &      &  &      &  &      &  &     &    &        &     &  31.60 &      &    \\
000813.9-343445 &    621& ESO349-031&   233& 14-13 13&  351.4707& -78.1179& 260.1831&   0.4018&  10& 0.012& 15.81&   229&  217&  222&  -40&       &  &      &      &  &      &  &      &  &     &    &        &     &        & 27.48& ~r~\\
000820.7-295458 &    627& NGC0007   &  1096& 61-18 18&   13.9903& -80.1369& 264.5891&   1.9321&   5& 0.014& 14.35&  1496& 1499& 1513& 1209&   0.21& 1&  0.00&  0.00& 0& 12.83& 1&  0.00& 0& 1496& 212&   31.81& 0.40&        &      &    \\
000956.4-245748 &    701& NGC0024   &   355& 19 ~-8  ~7&   43.6887& -80.4344& 269.3877&   3.2260&   5& 0.019& 12.10&   553&  572&  594&  249&   0.26& 2& 12.03& 11.05& 1& 10.56& 2&  9.75& 1&  553& 223&   29.79& 0.35&        &      &    \\
001124.7-412353 &    800& ESO293-045&  1088& 61  ~~0 16&  330.3100& -73.5343& 253.9490&  -2.4348&   8& 0.011& 15.25&  1466& 1430& 1423& 1229&   0.23& 1&  0.00&  0.00& 0& 14.20& 1&  0.00& 0& 1466& 156&   32.17& 0.40&        &      &    \\
001508.4-391313 &   1014& NGC0055   &   234& 14 13 13&  332.6677& -75.7388& 256.2418&  -2.4123&   9& 0.013&  8.47&   125&   95&   91& -121&   0.21& 1&  8.42&  7.57& 1&  7.21& 1&  0.00& 0&  125& 203&   25.98& 0.36&        & 26.70& ~r~\\
001531.5-321051 &   1038& ESO410-005&   234& 14 13 13&  357.8407& -80.7103& 262.9460&  -0.2577&  -5& 0.014& 15.17&     0&    0&    0&    0&       &  &      &      &  &      &  &      &  &     &    &        &     &        & 26.39& ~r~\\
001745.5+112701 &   1160& NGC0063   &  1213& 65  ~~6  ~6&  109.8744& -50.5655& 305.1594&  11.9146&   5& 0.111& 12.73&  1160& 1282& 1361&  805&       &  &      &      &  &      &  &      &  &     &    &        &     &        & 31.36& s~~\\
002023.1+591735 &   1305& IC0010    &   222& 14-12 12&  118.9699&  -3.3395& 354.4176&  17.8657&  10& 1.560& 11.78&  -346& -161&  -55& -565&       &  &      &      &  &      &  &      &  &     &    &        &     &        & 24.10& ~~c\\

\enddata
\end{deluxetable}

\begin{deluxetable}{rlrrrrrrrrrrrrrrrrrrrrrrrrr}
\tabletypesize{\scriptsize}
\rotate
\tablewidth{0pt}
\tablenum{2}
\tablecolumns{27}
\tablecaption{Averaged Distance Estimates for 743 Groups}
\label{tbl:dist_groups}

\tablehead{\colhead{Gp ID}&\colhead{NBG ID}&\colhead{$\ell$}&\colhead{$b$}&\colhead{SGL}&\colhead{SGB}&\colhead{log $M_B$}&\colhead{$V_{\odot}$}&\colhead{$V_{GSR}$}&\colhead{$V_{LS}$}&\colhead{$V_{CMB}$}&\colhead{$\mu$}&\colhead{$\epsilon_{\mu}$}&\colhead{$d$}&\colhead{SGX}&\colhead{SGY}&\colhead{SGZ}&\colhead{$V_{pec}$}&\colhead{N$_{LL}$}&\colhead{$\mu_{LL}$}&\colhead{$\epsilon_{LL}$}&\colhead{N$_{FG}$}&\colhead{$\mu_{FG}$}&\colhead{$\epsilon_{FG}$}&\colhead{N$_{src}$}&\colhead{$\mu_{src}$}&\colhead{$epsilon_{src}$}}

\startdata
   1& 11 -1  1&   282.93&  74.45&  102.70&  -2.35&  12.26&  1091& 1042&  999& 1421&   31.13& 0.10&   16.8&  -3.7&  16.4&  -0.7&  -246.&  51& 31.33& 0.11&  2& 30.92& 0.30& 91& 31.09& 0.10 \\
   2& 11  ~2  1&   299.61&  66.04&  112.68&  -1.26&  11.11&   900&  832&  776& 1235&   30.96& 0.13&   15.6&  -6.0&  14.4&  -0.3&  -376.&   9& 31.42& 0.16&  2& 31.45& 0.30&  2& 30.08& 0.17 \\
   3& 11 -3  1&   291.06&  68.94&  108.83&  -3.18&   9.77&  2011& 1946& 1894& 2349&   30.70& 0.36&   13.8&  -4.4&  13.0&  -0.8&   872.&   1& 30.70& 0.36&   &      &     &   &      &      \\
   4& 11 -4  1&   289.24&  65.45&  111.57&  -5.36&  11.12&  1615& 1538& 1480& 1960&   31.23& 0.11&   17.6&  -6.4&  16.3&  -1.6&   176.&   7& 31.47& 0.17&  1& 31.97& 0.41&  3& 30.94& 0.15 \\
   5& 11 -5  1&   283.70&  69.13&  107.33&  -5.36&  11.25&  1079& 1011&  960& 1421&   31.85& 0.11&   23.4&  -7.0&  22.3&  -2.2&  -775.&   7& 32.31& 0.16&  1& 31.69& 0.41&  5& 31.80& 0.11 \\
   8& 11  ~7  1&   245.25&  76.24&   94.22&  -6.75&   9.53&  1147& 1106& 1073& 1472&   29.64& 0.36&    8.5&  -0.6&   8.4&  -1.0&   446.&   1& 29.64& 0.36&   &      &     &   &      &      \\
   9& 11 -8  1&   299.14&  62.46&  116.08&  -2.64&  10.38&  1565& 1486& 1424& 1904&   30.98& 0.37&   15.7&  -6.9&  14.1&  -0.7&   262.&   1& 30.98& 0.37&   &      &     &   &      &      \\
  10& 11  ~9  1&   304.29&  62.04&  117.17&  -0.48&  10.23&  1233& 1158& 1096& 1566&   31.81& 0.15&   23.0& -10.5&  20.5&  -0.2&  -607.&   1& 31.27& 0.41&   &      &     &  1& 31.86& 0.16 \\
  15& 11  ~0  1&   283.19&  68.68&  107.56&  -5.78&  10.09&   748&  679&  626& 1091&   32.15& 0.28&   26.9&  -8.1&  25.5&  -2.7& -1366.&   2& 32.15& 0.28&   &      &     &   &      &      \\
  20& 11  ~0  1&   304.60&  78.71&  101.03&   3.72&   9.36&  1244& 1217& 1182& 1553&   30.87& 0.36&   14.9&  -2.8&  14.6&   1.0&    78.&   1& 30.87& 0.36&   &      &     &   &      &      \\
\enddata

\end{deluxetable}

\clearpage

\section{The Peculiar Velocity Field Within 3000 \kms}

Knowledge of distances, $d$, permits a subtraction of cosmic expansion velocities, H$_0 d$, from observed velocities, $V_{obs}$, to give $V_{pec}$, the radial component of what are referred to as peculiar velocities:
 \begin{equation}
V_{pec} = V_{obs} - {\rm H}_0 d
\end{equation}
where H$_0$ is the Hubble Constant.

The decomposition of observed velocities into cosmic expansion and peculiar velocity terms is seen to require knowledge of the Hubble Constant which is defined as
\begin{equation}
{\rm H}_0 = < V_{obs} / d>
\end{equation}
that is, a measure of the expansion rate over a sufficiently large domain of the Universe that peculiar motions cancel and have a negligible impact.

Imagine that observers make a zero point error in the determination of distances; i.e., on average, distances are off by a factor $f_e$ from true values, $d_{true} = f_e d_{measured}$.  Then the product H$_0 d$ has terms $f_e$ in the numerator and denominator that cancel.  The consequence is the well--known result that peculiar velocity measures are insensitive to a zero-point error in the distance scale as long as the assumed value of H$_0$ is consistent with the scale of measured distances.

Yet there is a problem.  We are not guaranteed that peculiar motions are negligible in the volume we sample to establish H$_0$.  For example, we live in the Local Supercluster which is an overdense part of the Universe.  It would not be surprising if there was a net infall within this region.  As a general statement, most observers in the Universe must live in overdense places, with a local retardation of the cosmic expansion, and will tend to measure a value of H$_0$ locally that is smaller than the cosmic value.  Or as another example, an observer might live on the outskirts of a large concentration and the preponderance of nearby galaxies in the direction of the concentration might be rushing away, toward the concentration.  The large number of these receding objects might cause H$_0$ to be overestimated.

In the present case, it is rather clear that the volume of our sample, limited to 3000 \kms, is too small to define H$_0$ without bias.   It might be tempting to assert that H$_0$ is known, for example from CMB measurements \citep{2003ApJS..148..175S}.  However such a value might not be consistent with the zero point scale of the distance measures.  Here we avoid the issue of which scale might be `correct', if there should be an inconsistency.  We simply note that H$_0$ on the scale of our present sample is not well defined because it does not extend in a self--consistent manner to large enough distances.   

These caveats regarding H$_0$ are mentioned because, as will be seen, there are large deviations from cosmic expansion seen within the 3000 \kms\ region {\it whatever} reasonable value is assumed for H$_0$.  Selecting a larger value of H$_0$ enhances a pattern in co-moving coordinates of overall infall while selecting a smaller value of H$_0$ creates a trend toward outflow.

The patterns of peculiar motion can best be seen in video animations posted at 

\noindent
ifa.hawaii.edu/$\sim$tully.

\subsection{The pattern of peculiar velocities and a choice of H$_0$}

With specification of H$_0$,  peculiar velocities can be found through Eq. (1) for all galaxies with measured distances.  Although there is uncertainty in H$_0$, observations constrain it to lie roughly within $70 < {\rm H}_0 < 80$ \kmsMpc.  The HST Key Project best estimate is toward the low end of this range \citep{2001ApJ...553...47F} while our own best estimate is toward the high side \citep{2000ApJ...533..744T}.

\begin{figure}[htbp]
\figurenum{2}
\centering
\includegraphics[scale=1.0]{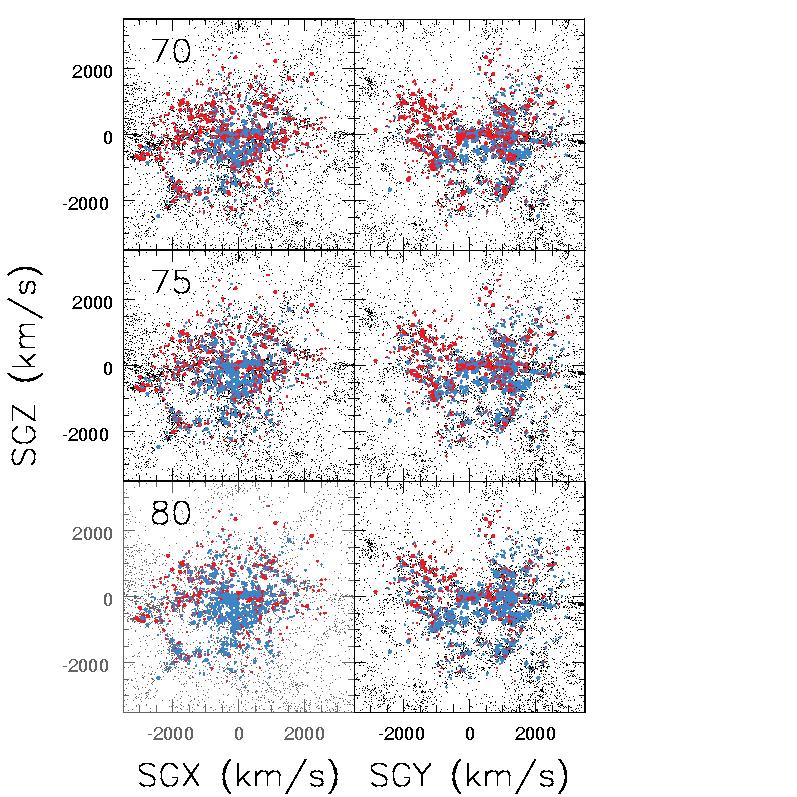}
\caption{Peculiar velocities from 2 views and with 3 choices for H$_0$.  {\bf Left:} SGX vs. SGZ in velocity units. {\bf Right:} SGY vs. SGZ. {\bf Top:} H$_0=70$~\kmsMpc. {\bf Middle:} H$_0=75$~\kmsMpc. {\bf Bottom:} H$_0=80$~\kmsMpc.  Large symbols: distances determined from Cepheids, TRGB, or SBF.  Small symbols: distances determined by the correlation between luminosity and linewidth.  Black dots: no distance available.  {\it Red}: peculiar velocities away from us.  {\it Blue}: peculiar velocities toward us.
}
\label{varyH}
\end{figure}

The series of panels in Figure~\ref{varyH} illustrates the effect of varying the choice of H$_0$ from 70, through 75, to 80 \kmsMpc.  Negative peculiar velocities are coded blue while positive peculiar velocities are red.  Large symbols are given in cases with Cepheid, TRGB, or SBF distances and small symbols are given in cases with the numerous but individually less accurate luminosity--linewidth distances.  The tiny black dots locate galaxies positioned according to their observed velocities but lacking distance measures.  There are 8795 galaxies in the $V<3500$~\kms\ cube,  part of a compilation drawn mainly from the Center for Astrophysics Redshift Catalog (http://www.cfa.harvard.edu/~huchra/zcat/zcom.htm) circa 2002.

\onecolumn
\twocolumn

Patterns of average positive and negative peculiar velocities can be seen in large swaths across these figures.  There is a prominent overall pattern of infall in the maps with H$_0=80$ \kmsMpc\ which
successively diminishes in the H$_0=75$ and H$_0=70$ maps.  These negative velocities dominate the map in a large sector toward the Virgo Cluster, the most populated region, and almost everywhere at negative SGZ;  i.e., below the equatorial plane in the supergalactic coordinate system.  By contrast, peculiar velocities  tend to be positive in the quadrant south of the Galactic plane (SGY negative) and above the supergalactic equator (SGZ positive).  Peculiar velocities also tend to swing positive at greater distances in the general direction of the motion indicated by the CMB dipole (near the supergalactic equator toward SGX negative).  The positive velocities are most pronounced in maps with H$_0=70$ but the trends persist with H$_0=75$ and $80$.

The major point we would make with this part of the discussion is that the overall patterns in the peculiar velocity field are similar whatever value for H$_0$ is considered in the range of reasonable values between 70 and 80 \kmsMpc.  The direction and amplitude of inferred peculiar velocity of our Galaxy is insensitive to the choice of H$_0$ over this range.  In Section 3.4, a weak preference will be found for H$_0=74$ \kmsMpc.  The amplitudes of peculiar velocities of individual galaxies other than our own depend on the choice of H$_0$.  The fundamental results of this paper are based on the well determined motion of our Galaxy and the patterns, but not critically the amplitudes, of other galaxies in our sample.

\subsection{Galactic and Local Group standards of rest} 

As a preliminary step, we review the status of the Solar motion with respect to the galaxies of the Local Group.  Here, as in the subsequent discussion, the amplitude and direction of our motion is determined by minimizing a condition of the following form:
\begin{equation}
{\rm min} [\sum_{i=1}^N ( V^i -{\rm H} d^i  +\hat x_i V_x +\hat  y_i V_y +\hat z_i V_z )^2]
\end{equation}
The $N$ galaxies to be considered with measured distances $d^i$ and observed velocities $V^i$ have Galactic coordinates $\ell_i$,$b_i$ which decompose along cardinal axes as
\begin{equation}
\hat x_i = {\rm cos} ~\ell_i ~{\rm cos}~ b_i 
\end{equation}
\begin{equation}
\hat y_i = {\rm sin} ~\ell_i ~{\rm cos} ~b_i 
\end{equation}
\begin{equation}
\hat z_i = {\rm sin} ~b_i 
\end{equation}
(or the equivalent $\hat  X$, $\hat Y$, $\hat Z$ in Supergalactic coordinates $L, B$).
One solves for the expansion component H and the cardinal components of our motion with respect to the chosen rest frame, $V_x, V_y, V_z$.  The term H can alternatively be fixed at a reasonable value or left as a free parameter.  In general, the solutions are more stable if H is fixed.

In the first step of the analysis of motions within the Local Group, heliocentric velocities are considered and the reference sample is $N=40$ galaxies within 1.1~Mpc, hence within roughly the zero--velocity surface or radius of first turnaround to infall in the Local Group \citep{2002A&A...389..812K}.  For this gravitationally bound sample the Hubble parameter is set to H$=0$~\kmsMpc.  The Sun is found to have a motion of $(V_x^2+V_y^2+V_z^2)^{1/2}=318 \pm 20$~\kms\ toward $\ell=106 \pm 4, b=-6 \pm 4$ ($L=349, B=+30$).  This solution is in good agreement with previous results   \citep{1977ApJ...217..903Y, 1996AJ....111..794K, 1999AJ....118..337C}.\footnote{Courteau (private communication) points out a misprint in his paper with van den Bergh. They intended to report $V_{LG}^{\odot}=306$~\kms\ toward $\ell=99, b=-3$.}
The close agreement with earlier work is expected since the  Local Group reference information has only been augmented incrementally.  It is instructive to note that the amplitude of 318~\kms\ is 12~km/s greater than found by  \citet{1999AJ....118..337C} because we include 14 additional galaxies, 5 of them dwarfs around M31 which turn out to have negative velocities larger than any previously known.  Though we now have 40 galaxies for the analysis they are strongly clustered on the sky and in distance.  If the sample is split between the 16 galaxies nearer than 500 kpc (the Milky Way companions) and the 24 more distant than 500 kpc (mostly the M31 sub-group) then the amplitude of the Solar motion with respect to these separate samples varies by $\pm 20$~\kms\ (342 and 299 \kms\ respectively).  The assigned error attempts to account for the effects of poor sampling.  Bootstrap resampling gives errors less than half what we quote.  The direction of the Sun's motion with respect to the Local Group has small errors because it is stabilized by the dominant component: the orbital motion of the Sun in the Galaxy.

Our solution provides the transform from the heliocentric rest frame, $V_{\odot}$, to the Local Group rest frame, $V_{LG}$ 
\begin{equation}
V_{LG} = V_{\odot} -86 \hat x +305 \hat y -33 \hat z
\end{equation}
 \begin{equation}
V_{LG} = V_{\odot} +270 \hat X -52 \hat Y +159 \hat Z
\end{equation}
The largest  component of this motion is due to the rotation of the Sun within the disk of the Milky Way. 
\citet{1997MNRAS.291..683F} 
claim the angular velocity at the Solar position is $27.19 \pm 0.87$~km~s$^{-1}$~kpc$^{-1}$ and review the details of the local motion of the Sun, while 
\citet{2005ApJ...628..246E} 
report a distance to the Galactic center of $R_0=7.62\pm0.32$~kpc.  The resultant transform of velocities from the Solar to the Galactic Standard of Rest is
\begin{equation}
V_{GSR} = V_{\odot} +9.3 \hat x +218 \hat y +7.6 \hat z
\end{equation}
corresponding to a motion of the Sun of $V_{GSR}^{\odot}=219\pm12$~\kms\ toward $\ell=87.6, b=+2.0$ ($L=356, B=+50$).
Then the motion of the Galaxy within the Local Group is
\begin{equation}
V_{LG}^{GSR}=V_{LG}-V_{GSR} = -95 \hat x +87 \hat y -41 \hat z
\end{equation}
or $135 \pm 25$~\kms\ toward {$\ell=137 \pm 10, b=-18 \pm 10$ ($L=342, B=-3$).  This direction is $17^{\circ}$ removed from the position of M31, offset toward the Maffei--IC~342 Group.  The projected positions of these features are shown in Figure~\ref{localmotion}.  The offset of the vector of our motion from M31 is sufficiently uncertain that a direct hit on M31 is not precluded.
\footnote{Throughout this discussion, subscripts on velocities will identify the reference frame and superscripts
will identify the object with the velocity. If there is no superscript it will be understood that the velocity pertains to object $i$ in an ensemble $N$.}

\begin{figure}[htbp]
\figurenum{3}
\centering
\plotone{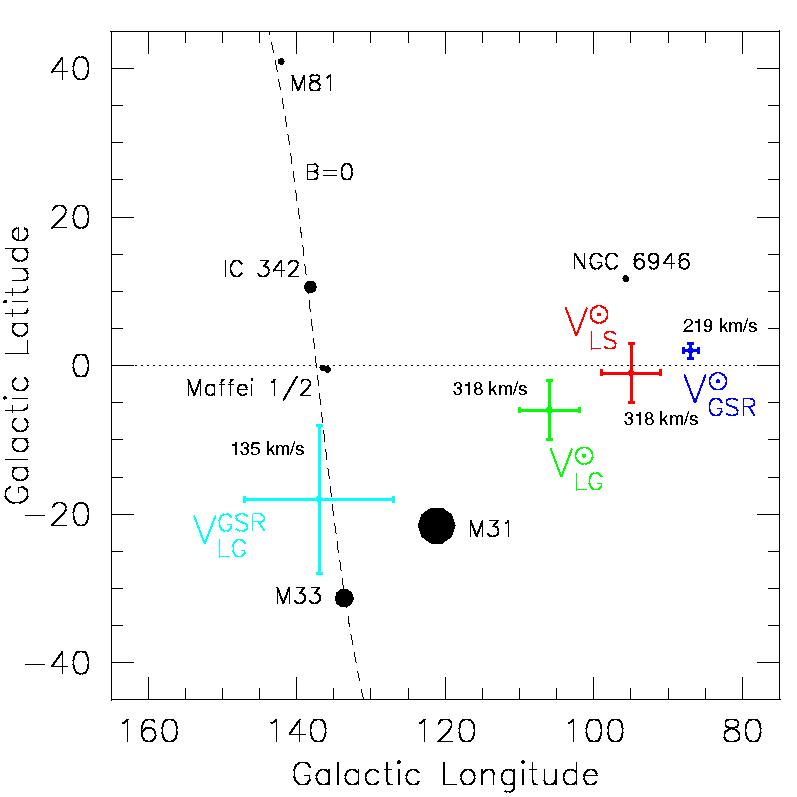}
\caption{The direction of solar motion.  The Sun has a motion $V_{GSR}^{\odot} = 219$~\kms\ about the center of our Galaxy and a motion $V_{LG}^{\odot} = 318$~\kms\ with respect to the centroid of the Local Group.  The vector difference $V_{LG}^{GSR}=V_{LG}-V_{GSR}=135$~\kms\ reflects the motion of the center of the Milky Way Galaxy in the Local Group rest frame. The other vector direction that is marked locates the motion $V_{LS}^{\odot} = 318$~\kms\ of the Sun with respect to galaxies beyond the Local Group but within 7~Mpc, within the Local Sheet. The dominant local attractor beyond the Galaxy is M31.  Ranked by apparent luminosity, and accepting luminosity as a stand--in for mass (both luminosity and gravitational attraction fall as $d^2$), M33 is an order of magnitude dynamically less important and IC~342 is a further factor of two less important.  The other galaxies identified in the figure are less important by further factors of 2 to  3.  The dashed line locates the Supergalactic equatorial plane.}
\label{localmotion}
\end{figure}

This Local Group rest frame may not deserve much attention.  Within this rest frame the Milky Way has a motion of 135~\kms\ essentially toward M31 while M31 has a motion of {\it zero} toward us.  With respect to the center of mass of the Local Group, in the absence of other forces, the Milky Way and M31 will have motions of approach that partition the observed 135~\kms\ in proportion to their masses.  The evidence from the motions of satellites suggest the two systems have comparable masses \citep{2000MNRAS.316..929E}.  Based on their relative luminosities, M31 could be expected to be 50\% more massive.  Certainly, the Milky Way is not a negligible test particle compared with M31.  The so-called Local Group rest frame is essentially the M31 rest frame.  In the next section we will investigate a more useful frame of reference.

\subsection{Peculiar velocities within the Local Sheet} 

It is at this second step that things get interesting.  Figure~\ref{local} zooms in from Figure~\ref{varyH}  (now for the case 
H$_0=74$~\kmsMpc) to highlight the local neighborhood. The color coding of velocities is more detailed.  We see a remarkable discontinuity in peculiar velocities between the galaxies that lie in our filament and the regions just beyond.  In the Nearby Galaxies Atlas \citep{1987nga..book.....T}, the structure we live in is called the Coma--Sculptor Cloud because it creates a band from Galactic north pole to Galactic south pole.  Our neighbors are tightly confined to the equatorial plane of the supergalactic coordinate system, so fall within the slice only $\pm 1.5$~Mpc thick about SGZ=0 shown in Figure~\ref{localsheet}.  The structure we live in and that has now been reasonably sampled with accurate distances has comparable dimensions in SGX and SGY.  This region is not quite synonymous with the Coma--Sculptor Cloud so we will refer to it as the `Local Sheet'.

The nearest adjacent structure lies in a layer at negative SGZ with respect to the Local Sheet and in the Nearby Galaxies Atlas is called the Leo Spur.  The abrupt step in peculiar velocities at the edge of the Local Sheet was called the `local anomaly' by \citet{1988lsmu.book..115F} and we called that step in conjunction with the apparent motion toward us of the Leo Spur the `local velocity anomaly'
\citep{1988lsmu.book..169T, 1992ApJS...80..479T}.   The large number of good TRGB distances available today place the local velocity anomaly in glaring relief.

The motion of the Local Group within the Local Sheet can be determined by realizing the minimization of Eq.~(7) for $N=158$ galaxies with measured distances in the range $1.1 < d_i<7$~Mpc; that is,  beyond those in the Local Group and nearer than those in the adjacent structures.  The best fit is found with H$=67$~\kmsMpc\ and a motion of the Local Group of $66 \pm 24$~\kms\ toward $L=150 \pm 37, B=+53 \pm 20$ ($\ell=349, b=+22$).  This motion differs from zero with only marginal significance and is consistent with the observation by \citet{2003A&A...398..479K} that nearby groups and individual isolated galaxies adhere to the local expansion with a dispersion of only 40~\kms.  Note that the local expansion need not be, and is probably not, the same as the cosmic expansion.

The very low relative peculiar velocities within the Local Sheet and our small, marginally significant peculiar motion within this structure suggest we consider a frame of reference that is at rest with respect to this structure.  The motion of the Sun with respect to 158 galaxies with accurate distances at $1.1 < d_i < 7$~Mpc in the Local Sheet defines the following relations:
\begin{equation}
V_{LS} = V_{\odot} -26 \hat x +317 \hat y -8 \hat z
\end{equation}
 \begin{equation}
V_{LS} = V_{\odot} +234 \hat X -31 \hat Y +214 \hat Z
\end{equation}
or a motion of the Sun of $V_{LS}^{\odot}=318 \pm 20$~\kms\ toward $\ell=95 \pm 4$, $b=-1 \pm 4$ ($L=353$, $B=+42$).\footnote{The best fit is found with the local expansion term H=67~\kmsMpc.  Rigorously, there should be a velocity correction to the centroid of the reference frame by the term $({\rm H}_0 - {\rm H})*d^{centroid} \sim 10$~\kms.}  The distance constraints are from Cepheid and TRGB methods and equally weighted.

This direction is shown in Fig.~\ref{localmotion}.  It is seen to be close in amplitude and direction with $V_{LG}^{\odot}$.  In fact, it is as close to our value of $V_{LG}^{\odot}$ as our value is to other estimates of the Local Group motion given in the literature.  Given both the uncertainties in $V_{LG}^{\odot}$ and the ambiguity in the meaning of this reference frame, we will base the rest of our discussion on the rest frame established by galaxies beyond the Local Group but within 7~Mpc, velocities we designate $V_{LS}$.  This reference frame is established by a completely independent sample than those that define $V_{LG}$ and is more robust, based on five times more galaxies and with good sky coverage.

\onecolumn
\figurenum{4}
\begin{figure}[htbp]
\centering
\includegraphics[scale=0.85]{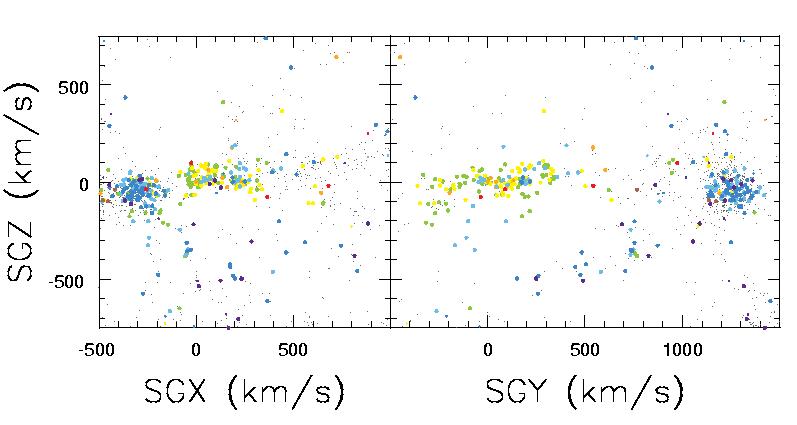}
\caption{Peculiar velocities in the local neighborhood. Zoom into the central region of the panels of Fig.~\ref{varyH}, now with H$_0=74$~\kmsMpc. In this figure and all the subsequent figures showing peculiar velocities, the color code is more refined than in Fig.~2.  Here: {\it purple} $V_{pec} \leq -400$~\kms; {\it dark blue} $-400 < V_{pec} \leq -200$~\kms; {\it light blue} $-200 < V_{pec} \leq -100$~\kms; {\it green} $-100 < V_{pec} < 0$~\kms; {\it yellow} $0 \leq V_{pec} < 100$~\kms; {\it orange} $100 \leq V_{pec} < 200$~\kms; {\it red} $200 \leq V_{pec} < 400$~\kms; {\it brown} $400 \leq V_{pec}$~\kms.  The left panel has a depth $-500<SGY<+1500$~\kms; the right panel: $-500<SGX<+1000$~\kms.}
\label{local}
\end{figure}
\twocolumn

\begin{figure}[htbp]
\figurenum{5}
\centering
\includegraphics[scale=0.56]{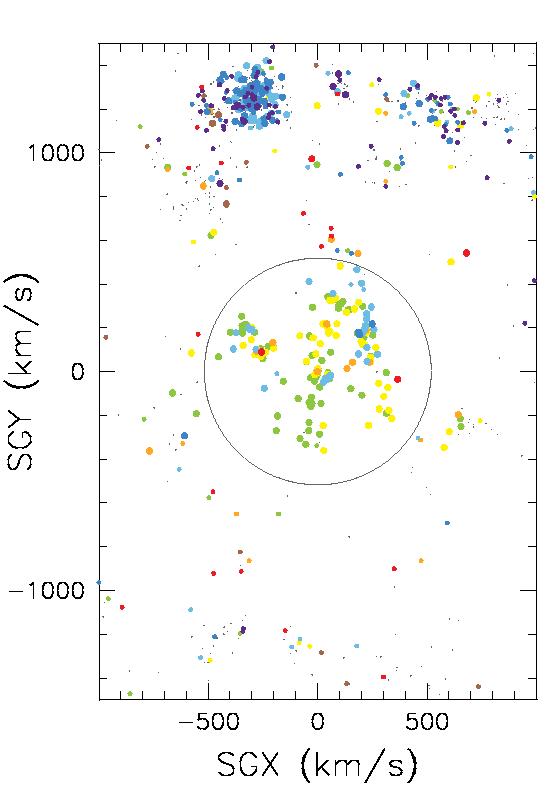}
\caption{
Peculiar velocities in and around the Local Sheet.  The region displayed is a slice 3 Mpc thick
centered on the supergalactic equator.  Symbols and colors have the same meaning as in previous plot (peculiar velocities are calculated assuming H$_0=74$~\kmsMpc).  The 7~Mpc radius circle centered on the Milky Way defines the region referred to as the Local Sheet.  The Virgo Cluster is at the upper left.
}
\label{localsheet}
\end{figure}

\subsection{The velocity discontinuity beyond the Local Sheet}

The availability of TRGB distances to objects beyond the Local Sheet have strongly confirmed the existence of a velocity discontinuity at $\sim 7$~Mpc.  The effect is unambiguously seen in the nearest part of the Leo Spur at large negative supergalactic latitudes.  The 14+19 association of dwarf galaxies       \citep{2006AJ....132..729T}  involves 4 galaxies with well established distances ($7.8 \pm 0.3$~Mpc) and velocities ($195 \pm 26$~\kms\ in the Local Sheet frame).  The derivation of peculiar velocities requires an assumption of the value of the Hubble Constant and, as will be justified later, we take
H$_0=74$~\kmsMpc.    In this case, members of the 14+19 association have peculiar velocities of $-382 \pm 47$~\kms.  In addition, two companions of NGC~2903, D564-08  and D565-06, have reliable TRGB distances of 8.4 and 8.5~Mpc and velocities of 385 and 394~\kms\ respectively which imply peculiar velocities of $-237$ and $-235$~\kms, and there is the extreme case of the relatively isolated Leo Spur galaxy D634-03 at 9.3~Mpc with $V_{LS}=186$~\kms\ and $V_{pec}=-502$~\kms\ \citep{2006AJ....131.1361K}.  These new high precision distances to galaxies with significant SGZ components in the line of sight confirm what had earlier been suspected: that the Leo Spur and our Local Sheet have peculiar motions of several hundred \kms\ toward each other.  The peculiar motions of these galaxies in the line-of-sight are indicated in Figure~\ref{leo}.

\begin{figure}[htbp]
\figurenum{6}
\centering
\plotone{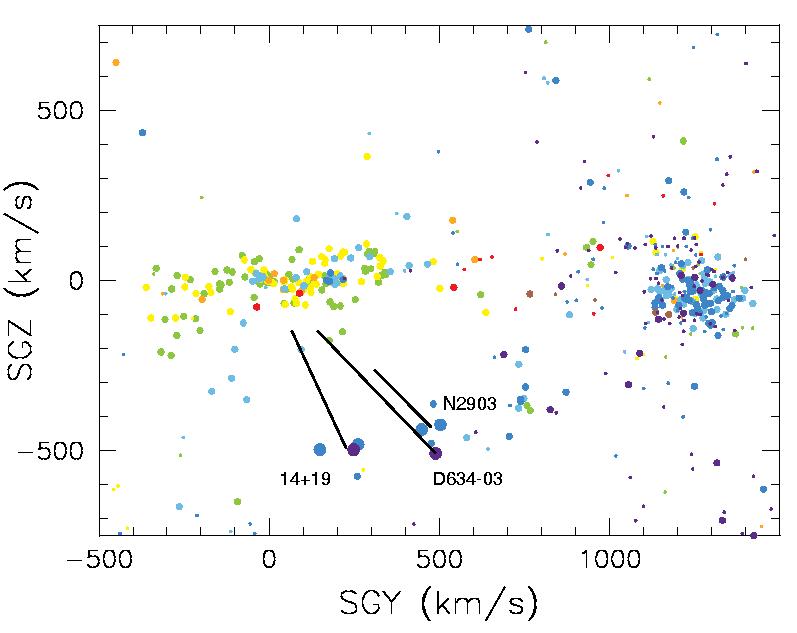}
\caption{Peculiar velocities in the local neighborhood.  Same projection as right panel of Fig.~\ref{local}.  Special attention is given to seven galaxies in three groups in the Leo Spur.
These galaxies have well determined TRGB distances that place them at $8-9$~Mpc and imply $V_{pec} \sim -325$~\kms.  The straight line vectors show the line-of-sight peculiar velocities averaged over each group assuming H$_0=74$~\kmsMpc.}
\label{leo}
\end{figure}

It is evident from the peculiar velocity patterns in Figure~\ref{varyH} that the anomaly is not restricted to the Leo Spur.  Negative peculiar velocities are seen all over in the region around the Virgo Cluster near the +SGY axis and generally at all $-$SGZ.  The negative peculiar velocities in the direction of the Virgo Cluster have long been seen as a reflex of the pull of the cluster on us \citep{1982ApJ...258...64A}.  This is part of the story but not all of it.  Another general feature is the trend of positive peculiar velocities in the quadrant with +SGZ and $-$SGY.

The occurrence of the velocity anomaly is manifested in an abrupt break in the amplitude and direction of galaxy motions relative to our motion found through the condition imposed by Eq.~(7).  The vector of our relative motion can be determined in shells of either distance or velocity to look for systematic drifts that would be indicative of the depth of perturbations or for erratic bounces that would indicate instability in the solution.   It is found that the vector of our motion is quite stable, achieving a direction and amplitude in the immediate shells beyond 7~Mpc that changes very little out to $V_{LS} \sim 3000$~\kms.  The global solution over this range with $N=683$ distance measures after averaging in groups provides the solution for transformation from Local Sheet referenced motions $V_{LS}$ to a reference frame established from objects within the general region of the Local Supercluster, $V_{LSC}$
\begin{equation}
V_{LSC} = V_{LS} -211 \hat x -178 \hat y +169 \hat z
\end{equation}
or
\begin{equation}
V_{LSC} = V_{LS} +35 \hat X +196 \hat Y -255 \hat Z .
\end{equation}
With respect to the general Local Supercluster reference frame, the Local Sheet has a motion of $V_{LSC}^{LS}=323 \pm 25$~\kms\ toward $\ell=220 \pm 7, b=+32 \pm 6$ ($L=80, B=-52$).  This best solution is achieved with H$_0=74$~\kmsMpc.  The solution is remarkably insensitive to the choice of H$_0$.  Variations from 60 to 90 \kmsMpc\ result in variations in the velocity amplitude of only $\pm 2$~\kms\ and variations in direction of only $\pm 5$ degrees.  With a choice of H$_0$ less than 74 there is an overall expansion and with H$_0$ greater than 74 there is an overall compression.  The value of H$_0=74$ is accepted for the rest of the discussion although it rests on the weak hypothesis that there is neither expansion nor compression centered on our location.  

Figure~\ref{aitoff_pecvel} provides a display of the currently available sample of peculiar velocities in an equal area projection on the sky.  Galaxies with $\vert V_{pec} \vert < 100$~\kms\ are not shown in order to make clear the separation on the sky between galaxies with large positive peculiar velocities and those with large negative values.  The crosses in the figure labeled LSC and CMB indicate vectors of motion that will be discussed in later sections.
In the following discussion, the volume beyond 7 Mpc and with $V_{LS}<3000$~\kms\ will frequently be referred to as simply the Local Supercluster (LSC).
 
 A significant component of the peculiar motion of the Local Sheet comes from the pull of matter in and near the Virgo Cluster. The cluster itself has a mass approaching $1 \times 10^{15}~\Msun$ \citep{2005ApJ...635L.113M}.  Numerical Action Method models demonstrate that this much mass in the cluster and a comparable amount of mass in the north Galactic hemisphere within the Local Supercluster generates a peculiar motion of $\sim 200$~\kms\ in the Virgo direction at our location -- as has long been implicated; eg \citet{1982ApJ...258...64A}.  The vector representing the motion of the Local Sheet with respect to the Local Supercluster, $V_{LSC}^{LS}$, has a component directed toward the Virgo Cluster of $V_{LSC;V}^{LS}=185 \pm20$~\kms.  If this vector toward Virgo is subtracted off the vector toward the overall Local Supercluster, the result is the vector $V_{LSC;LV}^{LS}$, where $LV$ stands for Local Void for reasons that will soon be described.  Coordinate frame transforms obey
 \begin{equation}
V_{LV} = V_{LS} -222 \hat x -130 \hat y -10 \hat z
\end{equation}
\begin{equation}
V_{LV} = V_{LS} +77 \hat X +16 \hat Y -248 \hat Z
\end{equation}
corresponding to a Local Sheet motion of $259 \pm 25$~\kms\ toward $\ell=210 \pm 7$, $b=-2 \pm 6$ ($L=11, B=-72$).  Since the Virgo and LV vectors are almost orthogonal, the decomposition has only a weak dependence on the amplitude of the Virgo component.  A variation of $\pm 50$~\kms\ in velocity toward Virgo affects $V_{LV}$ at the level of 10~\kms\ in amplitude and $15^{\circ}$ in direction.  The direction of the motion $V_{LSC}$ is shown in Figures~\ref{lsc_vectors_big} and \ref{lsc_vectors} along with the decomposition vectors $V_{LSC;V}^{LS}$ and $V_{LSC;LV}^{LS}$.

\onecolumn
\begin{figure}[htbp]
\figurenum{7}
\begin{center}
\includegraphics[scale=0.85]{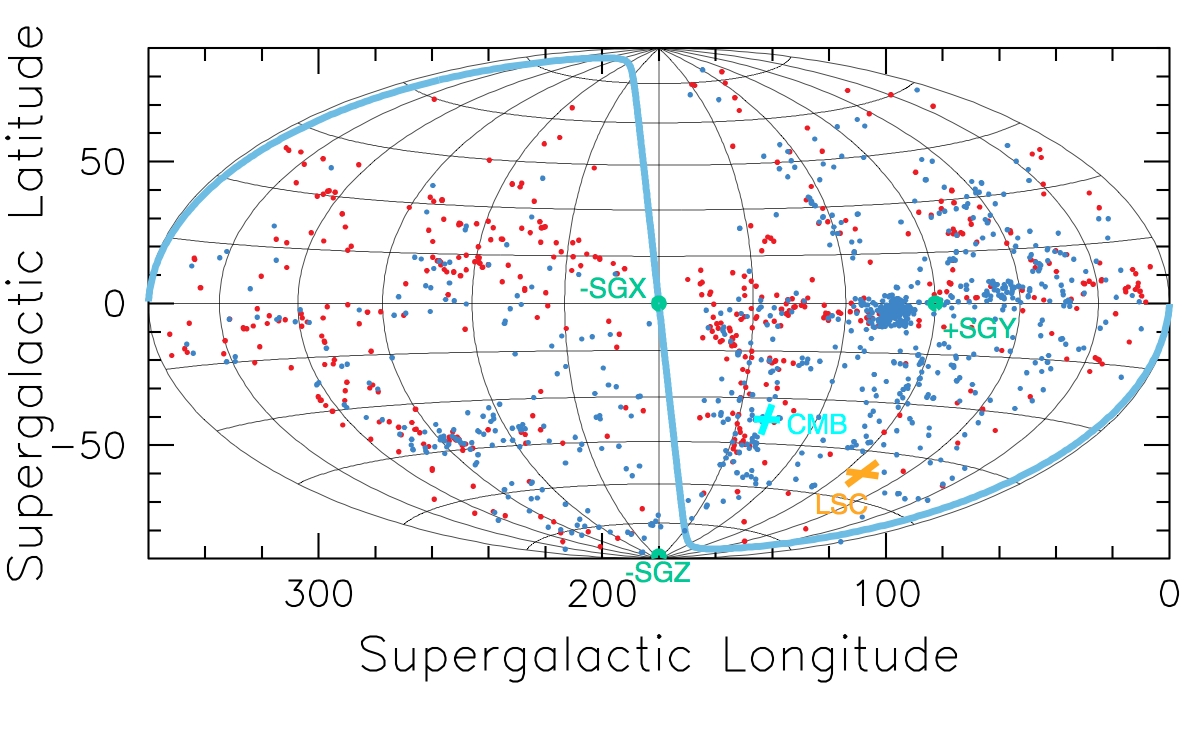}
\caption{Aitoff projection of observed peculiar velocities.  Blue symbols: $V_{pec}<-100$~\kms; red symbols: $V_{pec}>+100$~\kms.  The Local sheet has a motion with respect to this sample toward the orange cross labeled LSC and a motion toward the apex  of the Cosmic Microwave Background dipole at the position of the cyan cross labeled CMB.  The heavy blue line defines the plane of our Galaxy.  The knot of blue symbols at $L=103$, $B=-2$ is the Virgo Cluster.}
\label{aitoff_pecvel}
\end{center}
\end{figure}
\twocolumn

\begin{figure}[htbp]
\figurenum{8}
\begin{center}
\plotone{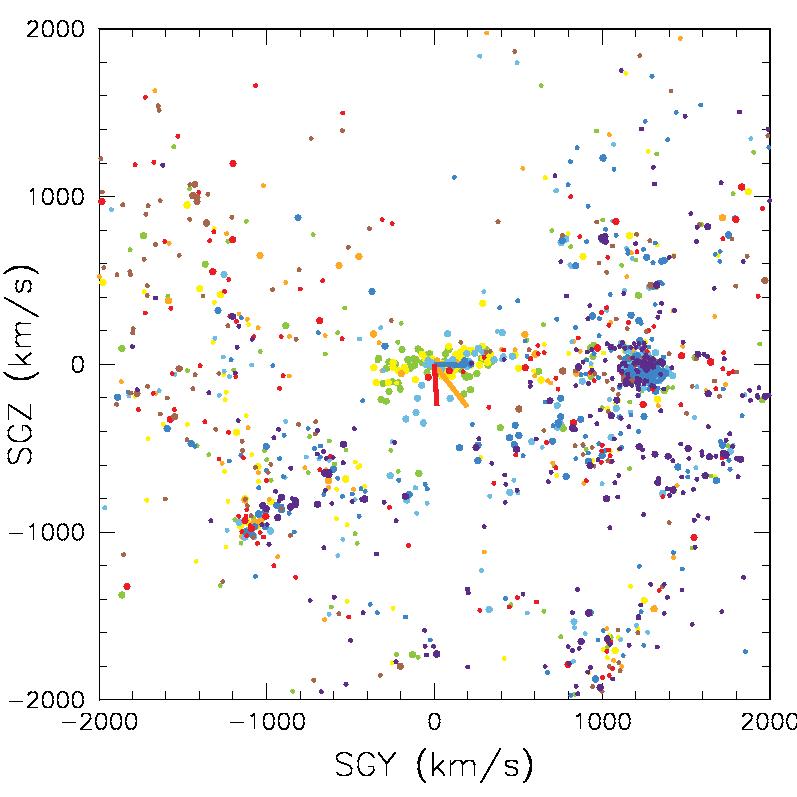}
\caption{Motion within the Local Supercluster in the rest frame of the Local Group.  Galaxies with $-2000<SGX<2000$~\kms\ are plotted.  Peculiar velocities are color coded as in previous figures.  The vectors emanating from our position at the origin indicate our motion relative to these galaxies.  They are described in the following figure which is an enlargement of the central region of this figure.}
\label{lsc_vectors_big}
\end{center}
\end{figure}

\begin{figure}[htbp]
\figurenum{9}
\centering
\plotone{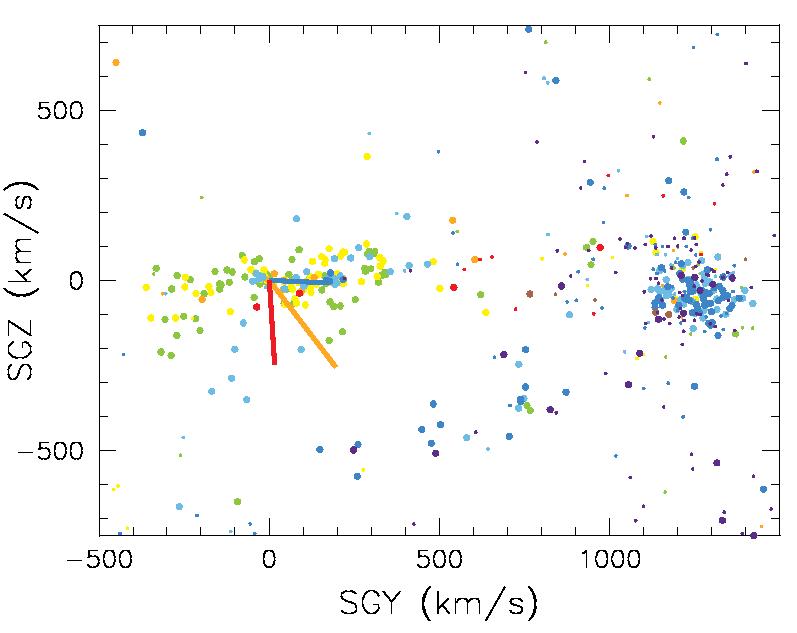}
\caption{Motion within and around our home structure, the Local Sheet, with $-500<SGX<1000$~\kms.  The orange vector represents $V_{LSC}^{LS}$ with an amplitude of 323~\kms\ in the rest frame of the Local Sheet.  The blue vector has an amplitude of 185~\kms\ and is directed toward the Virgo Cluster at the right edge of the figure.  The red vector is the residual of these two, called $V_{LSC;LV}^{LS}$, and has an amplitude  of 259~\kms.}
\label{lsc_vectors}
\end{figure}

\subsection{The Local Void}

The vector defined by Eqs.~(19,20) is not pointing at anything prominent but it is directed {\it away} from the Local Void.  This negative feature was identified in the Nearby Galaxies Atlas.  The possible influence of the Local Void has been anticipated   \citep{1988lsmu.book..115F, 1988MNRAS.234..677L}.  There is the claim that the far side of the void is in expansion away from us  \citep{2005ASPC..329...59I}.  The significance of the Local Void has been difficult to evaluate because it is intersected by the zone of obscuration but the neutral Hydrogen survey HIPASS substantiates its importance \citep{2004MNRAS.350.1195M}.

Figure~\ref{void} attempts a visualization of the Local Void.  This absence of galaxies begins at the edge of the Local Group at positive SGZ.  It appears to consist of a void within larger voids; i.e., a smaller void shares an interior wall of a larger almost empty region.  We lie on a filament that serves as a wall for both the smaller and larger voids.  Even the smaller void is not so small, with a long dimension of  $\sim 35$~Mpc.  The geometry of the larger enclosing void is quite uncertain.  It appears to be bisected by a filament into north and south parts.  The long dimension may be as large as 5,000~\kms\ $\sim 70$~Mpc.  In the entire region, but especially with the larger component, aspects of the voids are poorly defined because of interruption by the zone of obscuration (roughly coincident with SGY=0).  The near and split far underdense regions will be referred to as the Inner, North, and South Local Voids, or in the ensemble as just the Local Void.

Motions on the far walls of the Local Void are poorly documented because of their distance and problems caused by obscuration.  
Current distance estimates for galaxies at 25--30~Mpc bounding the Local Void have peculiar velocities $\sim +300$~\kms.
For the moment, these distances do not have sufficient quality to distinguish peculiar motions at the far wall of the Local Void from the reflex of the motion of the Local Sheet. 

\onecolumn
\begin{figure}[htbp]
\figurenum{10}
\centering
\includegraphics[scale=0.85]{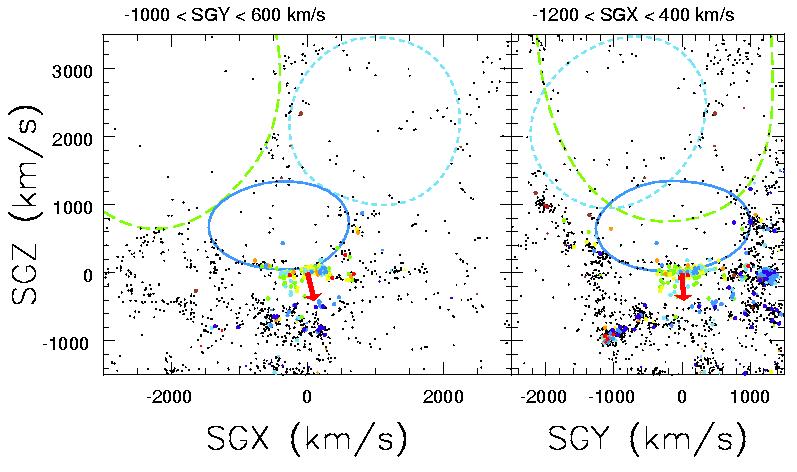}
\caption{The region of the Local Void.  The ellipses outline the 3 apparent sectors of the Local Void. The solid dark blue ellipses show two projections of the Inner Local Void bounded on one edge by the Local Sheet.  The North and South extensions of the Local Void are identified by the light blue short-dashed ellipses and the green long-dashed ellipses, respectively.  These separate sectors are separated by bridges of wispy filaments.
The red vector indicates the direction and amplitude of our motion away from the void.}
\label{void}
\end{figure}
\twocolumn

The Local Void being a void, there is not much opportunity to measure motions {\it within} the void, but we are offered at least one chance.  An HST observation provides a TRGB measurement for the lonely galaxy ESO~461-36 = KK~246 \citep{2006AJ....131.1361K}.  The distance given in that reference is probably too great, primarily because the reddening estimate that was used  \citep{1998ApJ...500..525S} is too low.  Using the procedures described by  \citet{2007ApJ...661..815R}   we find a distance of 6.4~Mpc.  Though closer than previously suspected, the galaxy still lies well into the Local Void.  This galaxy has an observed $V_{LS}=443$~\kms\ resulting in $V_{pec}=-30$~\kms\ with H$_0=74$.  However, ESO~461-36 is at almost the opposite pole from the Local Sheet motion described by Eqs.~(17,18).  Its motion with respect to the Local Supercluster is roughly the sum of our motion and its additional motion in the same direction (discounting proper motion components).  Hence this galaxy is trying to escape from the void with a deviant velocity of at least 350~\kms.  The situation is seen in Figure~\ref{lvdw}.  ESO~461-36 has a peculiar velocity toward us in the Local Sheet rest frame as do galaxies on almost the opposite side of the sky in the Leo Spur.  However in the rest frame established by galaxies with known distances in the Local Supercluster we are moving toward the Leo Spur and away from ESO~461-36.  With respect to the LSC, ESO~461-36 has a very high peculiar velocity.

\begin{figure}[htbp]
\figurenum{11}
\begin{center}
\plotone{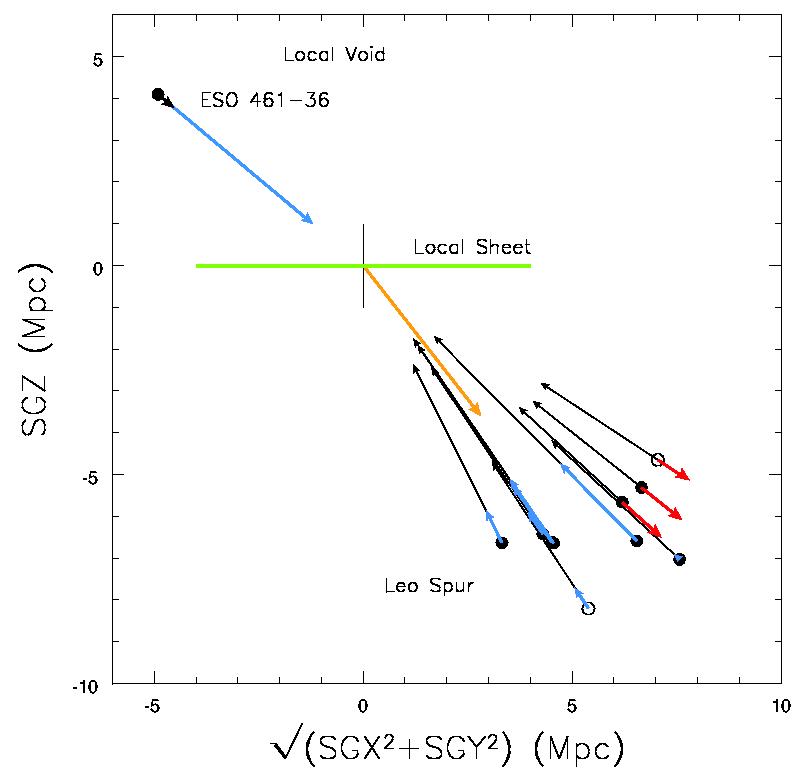}
\caption{Motions of galaxies with accurate distances in the Leo Spur and Local Void.  Horizontal axis: vector sum of SGX and SGY components of distance.  Vertical axis: SGZ component of distance.  Filled circles: TRGB distances.  Open circles: SBF distances.  Black vectors: peculiar velocity assuming H$_0=74$~\kmsMpc.  Orange vector: motion of Local Sheet with respect to galaxies with measured distances within the Local Supercluster ($V_{LS}<3000$~\kms\ sample), $V_{LSC}^{LS}=323$~\kms\ toward $L=80, B=-52$.  The blue and red vectors are the residuals of the black vectors after vector addition of the component of the orange vector in their lines of sight (blue: residual toward our position; red: residual away from our position).  In the case of the isolated galaxy ESO 461-36 in the Local Void the components add to a velocity of 349~\kms\ toward us in the Local Supercluster reference frame.}
\label{lvdw}
\end{center}
\end{figure}

Figure~\ref{lvdw} provides more details than Fig.~\ref{leo} concerning the motions of galaxies below the supergalactic equatorial plane.  All the galaxies indicated in the plot lie in the Leo Spur and have accurately known distances and systemic velocities.  Assuming H$_0=74$~\kmsMpc,
all galaxies in this sector have substantial peculiar velocities toward us.  The average motion for the 10 good cases in the figure is $-335$~\kms\ in the Local Sheet rest frame.  Upon cancelation of the motion of the Local Sheet with respect to the Local Supercluster , the average residual for these 10 galaxies is $-34$~\kms\ with a standard deviation of 29~\kms.  To within the errors, velocities in the Leo Spur can be viewed as simply the reflex of our motion in that direction.  We should be reminded, though, of the continuing uncertainty in the parameter H$_0$.  The average residual for these 10 galaxies would be nil if the assumed value of the Hubble Constant is reduced by $\Delta {\rm H} = -3$.  By the same token,  the residual would be significant if $\Delta {\rm H} >+3$.

\begin{figure}[htbp]
\figurenum{12}
\begin{center}
\plotone{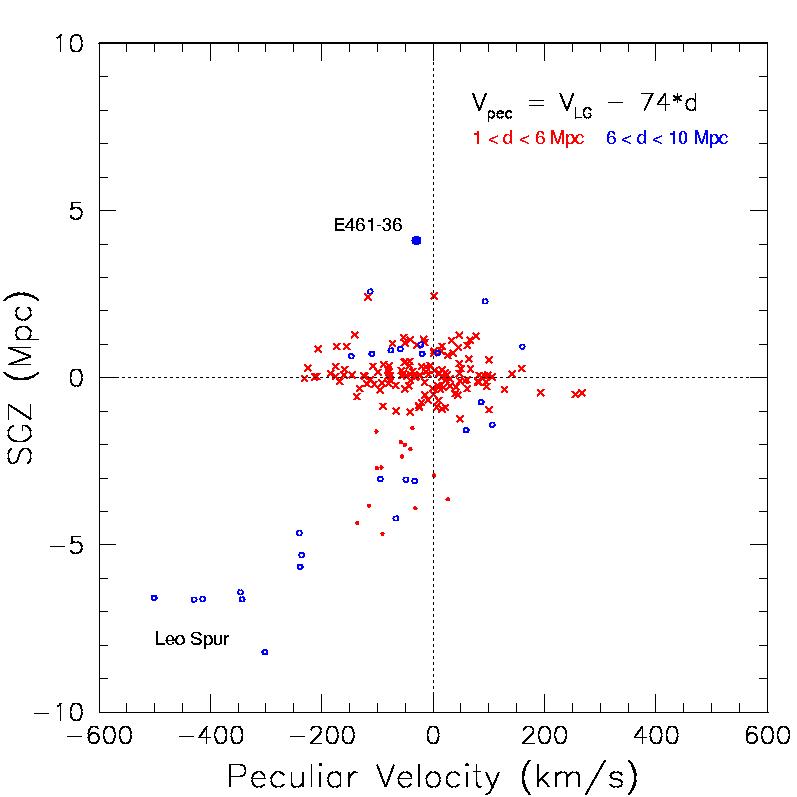}
\caption{Peculiar velocity as a function of distance SGZ from the supergalactic plane.  Red symbols:
distances of 1 to 6 Mpc; crosses: galaxies in the equatorial plane, and small dots: galaxies in a
flare to negative SGZ in the vicinity of $\ell \sim 270$.  Blue circles: distances between 6 and 10 Mpc.  The galaxies with the most deviant negative peculiar velocities at the most negative values of SGZ lie within the Leo Spur.  The galaxy in the Local Void, ESO 461--36, is identified by the large filled circle.
}
\label{sgz}
\end{center}
\end{figure}

Returning nearer to home, we can ask if there is a gradient of peculiar velocity with SGZ {\it within} the Local Sheet.  We look for this possibility with Figure~\ref{sgz}.  The distance from the supergalactic equatorial plane is plotted against peculiar velocity.  Galaxies inside and outside the filament are distinguished by color and symbol shape.  The general trend  of negative velocities can be interpreted as an overall local retardation from the mean cosmic expansion.  The largest negative $V_{pec}$ are seen in the Leo Spur.  Restricting attention to the galaxies within $\pm 1.5$~Mpc of the plane of our filament, one finds a marginal offset in peculiar velocities between positive and negative SGZ: $<V_{pec}>^{+SGZ} = -33 \pm 10$~\kms\ for 80 cases and $<V_{pec}>^{-SGZ} = 0 \pm 13$~\kms\ for 54 cases, a difference of $33 \pm 16$~\kms.  

The flare of galaxies at $-5<SGZ<-1.5$ Mpc off the Local Sheet seen in Fig.~\ref{sgz} is a minor feature that includes NGC~1313 and intrinsically smaller galaxies.  For 14 cases, $<V_{pec}>^{flare} = -63 \pm 12$~\kms.  These galaxies are moving toward  positive SGZ with respect to the Local Sheet. However in the LSC frame they are moving toward negative SGZ, like us but not as rapidly.

\subsection{The large--scale component of our peculiar velocity}

The Local Supercluster motion expressed by Eqs.~(17,18) fails in both amplitude and direction to explain the motion indicated by the cosmic microwave background.  The principal sources of that motion are suspected to lie at distances in velocity of 3,000 $-$ 6,000~\kms\ \citep{1988ApJ...326...19L,2006MNRAS.368.1515E} if not out at 10,000 $-$ 15,000~\kms\ \citep{1989Natur.338..562S, 2006ApJ...645.1043K}.  The sample of distances used in this paper reaches only to the inner edge of the nearer of these domains.  Perturbations consistent with the large--scale flows discussed by others  \citep{2000ApJ...530..625T}  are seen at the edge of our field of study at large $-$SGX.  The Centaurus Cluster with $V_{LS}^{Cen} = 3152$~\kms\ at $d^{Cen} = 37$~Mpc has $V_{pec}^{Cen} = +429$~\kms\ if H$_0 = 74$~\kmsMpc.

To a first approximation, the local and large-scale components of our motion can be treated as decoupled.   Let us determine the properties of the large--scale component upon subtraction of the local component from the CMB vector.  The transform between our Local Sheet frame and the reference frame of the CMB \citep{1996ApJ...473..576F} is given by 
\begin{equation}
V_{CMB} = V_{LS} +1 \hat x -563 \hat y +285 \hat z
\end{equation}
\begin{equation}
V_{CMB} = V_{LS} -381 \hat X +331 \hat Y -380 \hat Z
\end{equation}
which describes a motion of the Local Sheet of $631 \pm 20$~\kms\ toward $\ell = 270 \pm 3, b = +27 \pm 3$ ($L=139, B=-37$).  Subtraction of the Local Supercluster motion of Eqs.~(17,18) from the CMB motion:
\begin{equation}
V_{CMB} - V_{LSC} = +212 \hat x -385 \hat y +116 \hat z
\end{equation}
\begin{equation}
V_{CMB} - V_{LSC} = -416 \hat X +135 \hat Y -125 \hat Z
\end{equation}
describes a motion of $455 \pm 15$~\kms\ toward $\ell = 299 \pm 3, b = +15 \pm 3$ ($L=162, B=-16$).  The vector of motion of the Local Sheet indicated by the CMB dipole, $V_{CMB}^{LS}$, and the residual to this vector after the locally generated component $V_{LSC}^{LS}$ is subtracted off are shown in Figure~\ref{allv}. The direction of this large--scale component is closely aligned with the Norma--Hydra--Centaurus  supercluster complex and background Shapley Concentration, lying within $7^{\circ}$ of the direction of the Centaurus Cluster.

There is a recapitulation of the various reference frames and vectors in Table~3, and Figure~\ref{aitoffv} provides a visual summary.    The projected locations of the various vectors are indicated on this plot.   The CMB vector can be decomposed into the vector determined by motions within 3000~\kms\  (the Local Supercluster component) and a residual attributed to structure on large scales.  The Local Supercluster component can be decomposed in turn into the components toward the Virgo Cluster and away from the Local Void. 

It is known that the distribution of various populations of galaxies peak in roughly the direction of the CMB dipole maximum.  Two recent studies are considered here.  \citet{2006MNRAS.368.1515E} have calculated the dipole in the distribution of  sources brighter than $K_s=11.25$ from the Two-Micron All-Sky Redshift Survey (2MRS).  \citet{2006ApJ...645.1043K} have made the equivalent determination based on the distribution of X-ray selected clusters of galaxies.   These alternatively derived dipole directions are plotted in Fig,~\ref{aitoffv}.   In both cases, these dipole directions are within 20 degrees of the CMB dipole direction.  However, they are offset in revealingly different directions.  The 2MRS dipole is offset toward the Local Supercluster component of our motion and the X-ray dipole is offset toward the large scale component of our motion.

These distinctive offsets can be understood by giving consideration to Figure~\ref{dipoles} adapted from \citet{2006ApJ...645.1043K}.  The bottom panel gives histograms of the run of sources with velocity in the 2MRS and X-ray samples.  The 2MRS sample peaks at 5,000~\kms\ while the X-ray sample peaks at 18,000~\kms.  In the top panel we see that the 2MRS dipole amplitude crests at 4,000~\kms\ and then is flat  \citep{2006MNRAS.368.1515E}.  By contrast, the X-ray dipole peaks twice, once at 4,000~\kms\ and then again around 20,000~\kms\  \citep{2006ApJ...645.1043K}.  The shot noise is greater with the X-ray sample.  However, it is rather convincing that the two dipole investigations are sensitive to two separate features in the distribution of matter.  The 2MRS dipole is strongly influenced by nearby structure and insensitive to structure beyond 13,000~\kms.  We can appreciate why the dipole in the objects mapped with the 2MRS is pulled from the CMB direction toward the vector of Local Supercluster motion.  By contrast, the X-ray dipole only starts to build at 3,000~\kms\ so is strictly a reflection of the distribution of matter on large scales.  It is not a surprise that the direction of the X-ray dipole is pulled from the CMB direction toward the direction of the large scale component of our motion.

The 2MRS sample is attractive because redshifts are available for almost all the galaxies.  This information can be used to construct dynamical models \citep{2006MNRAS.373...45E}.  However if mass is distributed like light then, since both luminosity and gravity diminish as the square of distance, the net attraction on the Galaxy should be described by the full, deep 2MASS sample, without recourse to redshifts.  The analysis by \citet{2003ApJ...598L...1M} produced a dipole that moves from the 2MRS position $22^{\circ}$ W of the CMB position to $8^{\circ}$ W in Galactic longitude.  The Maller et  al. dipole position is $15^{\circ}$ N of the CMB position in Galactic latitude but this displacement may be due to the way the mask of the obscured region of the Galactic plane is filled.  This is the region of the Local Void.  If the region of the Local Void were given far fewer sources in the Maller et al. mask, the full 2MASS dipole would be pushed close to the CMB target.  It can be noted that the 2MRS analysis uses more information at low Galactic latitudes and gets a closer fit to the CMB in latitude (though it was mentioned that this shallower survey gets a worse fit in longitude).

\onecolumn
\begin{figure}[htbp]
\figurenum{13}
\begin{center}
\includegraphics[scale=0.8]{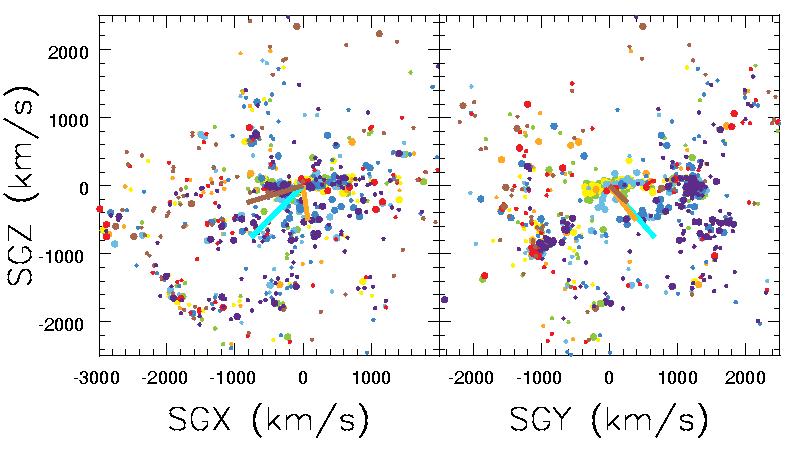}
\caption{Decomposition of the vectors of the motion of the Local Sheet.  Two orthogonal views are shown and peculiar velocities of galaxies with observed distances are shown with the same color code introduced in Fig.~\ref{local}.  The orange vector indicates the motion of the Local Sheet with respect to this sample.  The blue vector indicates the motion of the Local Sheet with respect to the rest frame established by the CMB.  The brown vector is the vector difference between these two and is attributed to influences on scales greater than 3,000~\kms.  For clarity, the lengths of these three vectors are doubled compared with the scales on the axes.}
\label{allv}
\end{center}
\end{figure}

\begin{figure}[htbp]
\figurenum{14}
\begin{center}
\includegraphics[scale=0.85]{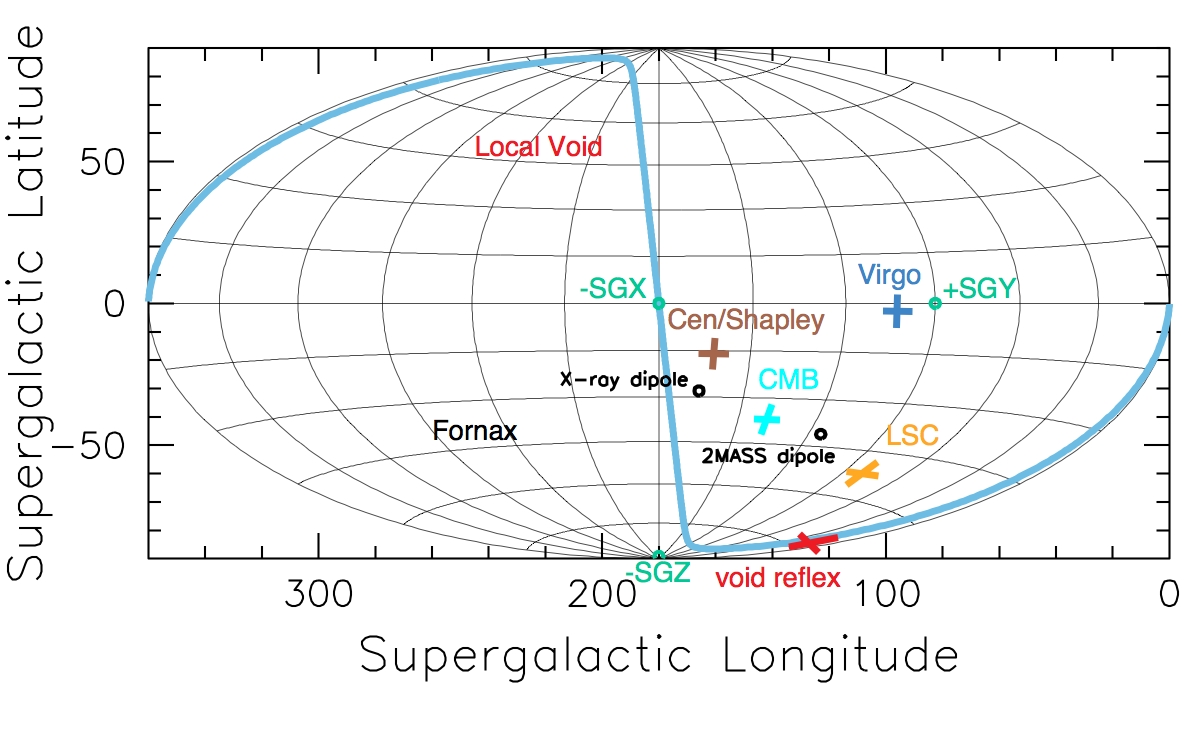}
\caption{Components of the motion of the Local Sheet projected onto the sky.  The motion of 631~\kms\ given by the CMB dipole in the direction indicated by the cyan cross can be decomposed into the 
323~\kms\ component defined by the distance measures discussed in this paper, confined to the traditional Local Supercluster, and labeled LSC in orange and the 455~\kms\ residual located by the brown cross that can be ascribed to large scale structures in Norma--Hydra--Centaurus and to the background.  The  Local Supercluster $<3000$~\kms\ component can in turn be separated into a component of 185~\kms\ toward the Virgo Cluster, at the blue cross, and a component of 259~\kms\ away from the Local Void, at the (severely distorted) red cross.  The Local Void reflex, Virgo, and large scale Cen/Shapley attractions are almost orthogonal to one another, toward the --SGZ, +SGY, and --SGX axes respectively.  The X-ray dipole direction is found to lie close to the direction of the large scale Cen/Shapley vector which can be understood since the characteristic distances of the X-ray cluster sample are large.  By contrast the 2MASS  dipole from the 2MRS lies midway between the CMB and Local Supercluster vector directions.  It is inferred that the 2MRS dipole is determined relatively locally.}
\label{aitoffv}
\end{center}
\end{figure}
\twocolumn

\begin{figure}[htbp]
\figurenum{15}
\begin{center}
\includegraphics[scale=0.5]{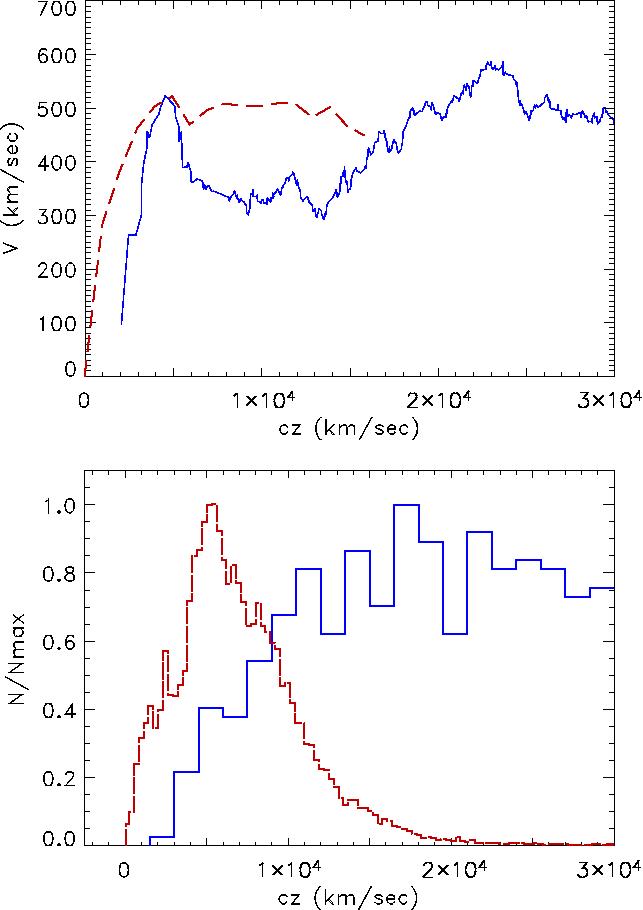}
\caption{X-ray and near infrared dipole amplitudes.  Top panel: The solid blue line shows the
development of the number-weighted X-ray cluster dipole amplitude with redshift.  The dashed red line shows the equivalent information for 2 micron selected sources.  Bottom panel: Histograms of the 
redshift distributions of the X-ray and 2 micron selected samples.
}
\label{dipoles}
\end{center}
\end{figure}

\section{Discussion}

In the future, the database of galaxy distances and peculiar velocities will be used for detailed studies of the distribution of matter using non-parametric Numerical Action Methods \citep{1995ApJ...454...15S}, techniques that can be used on small scales and at high densities.   For the moment, the discussion is restricted to first order effects.  It has been emphasized that the motion of the Local Sheet reflected in the CMB dipole can be decomposed into three main components.  Of course this is a simplification and taken to the other extreme of complexity this motion can be broken into an arbitrarily large number of influences.  

The particular interest in this study is the influence of the Local Void on our motion.  First, though, a few words are in order regarding the other two principal components.   Concerning scales larger than 3000~\kms, we would only point out here that a larger local contribution to the CMB motion implies a smaller value of $\beta = \Omega_m^{0.6} / b$ as derived from the amplitude of the dipole of large scale features.  Here, $\Omega_m$ is the mean density of matter compared with the density of matter that would give a closed universe and $b$ is the bias between the distribution of matter and the distribution of observable tracers.   For example, the values of $\beta$ calculated by   \citet{2006ApJ...645.1043K} from the X-ray cluster sample, which has the dependence $\beta = V_{pec} / D_{cl}$, where $D_{cl}$ is the dipole amplitude found from the X-ray clusters, is reduced by the lower large scale component of $V_{pec}$ found here by 11\% from the values given by Kocevski \& Ebeling.  

The component of our motion attributed to infall toward the Virgo Cluster was discussed by \citet{2005ApJ...635L.113M}.  This component is particularly amenable to modeling by Numerical Action Methods with the large number of distance constraints that are becoming available.  We reserve further discussion for another paper but emphasize the relative decoupling from the motion away from the Local Void because (a) the two components are almost orthogonal, and (b) the scale of the Virgo Cluster influence is governed by the cluster distance of 17~Mpc while there are sharp gradients attributed to Local Void  effects on scales of only a few Mpc.

\subsection{Expansion of Voids}

We turn now to consider the reflex motion from the Local Void.  First, what can be expected on theoretical grounds? 
The Friedmann Equation can be written as:
\begin{equation}
 {\rm H}^2 \equiv \left(\frac{\dot{a}}{a}\right)^2 = \frac{8 \pi G \rho}{3} + \frac{\Lambda}{
3} - K\frac{c^2}{a^2} .
\end{equation}
where $a$ is the radial scale factor normalized to $a=1$ today and the three terms on the right describe contributions from the mean density of matter, $\rho$, the vacuum energy, $\Lambda$, and spatial curvature, $K$. 
Within a completely empty void:
\begin{equation}
\dot{a}^2 = (\Lambda/3) a^2 - K c^2 .
\end{equation}
This expression can be related to global parameters with $\alpha = {\rm H_v / H}_0$ where ${\rm H_v} = \dot{a}/a$ inside the void and $\Omega_{\Lambda} = \Lambda / 3 {\rm H}_0^2$.
\begin{equation}
\dot{a}={\rm H_v} \left(\frac{\Lambda a^2}{3 {\rm H_v}^2} - \frac{K c^2}{{\rm H_v}^2} \right)^{1/2}
\end{equation}
then since the curvature term in the void is 
\begin{equation}
\frac{-K c^2}{{\rm H_v}^2} = 1 - \frac{\Lambda}{3 {\rm H_v}^2}
\end{equation}
we arrive at
\begin{equation}
\dot{a}=\alpha {\rm H}_0 \left(\frac{\Omega_{\Lambda} a^2}{\alpha^2} +(1- \frac{\Omega_{\Lambda}}{\alpha^2}) \right)^{1/2} .
\end{equation}
We solve for the value of $\alpha$ that takes $a$ from 0 to 1 in time $t_0$ for the case $\Omega_m=0.24$, $\Omega_{\Lambda}=0.76$, H$_0=74$~\kmsMpc, and $t_0=13.7$~Gyr.  We find $\alpha=1.22$, which gives an expansion in the void relative to the universal flow of 
\begin{equation}
{\rm H_v - H}_0 = 16 ~{\rm km~s^{-1}~Mpc^{-1}}. 
\end{equation}
 For comparison, with $\Omega_m=0.3$, $\Omega_{\Lambda}=0.7$ one finds $\alpha=1.25$ and  ${\rm H_v - H}_0 = 18$~\kms.

\begin{figure}[htbp]
\figurenum{16}
\begin{center}
\includegraphics[scale=0.4]{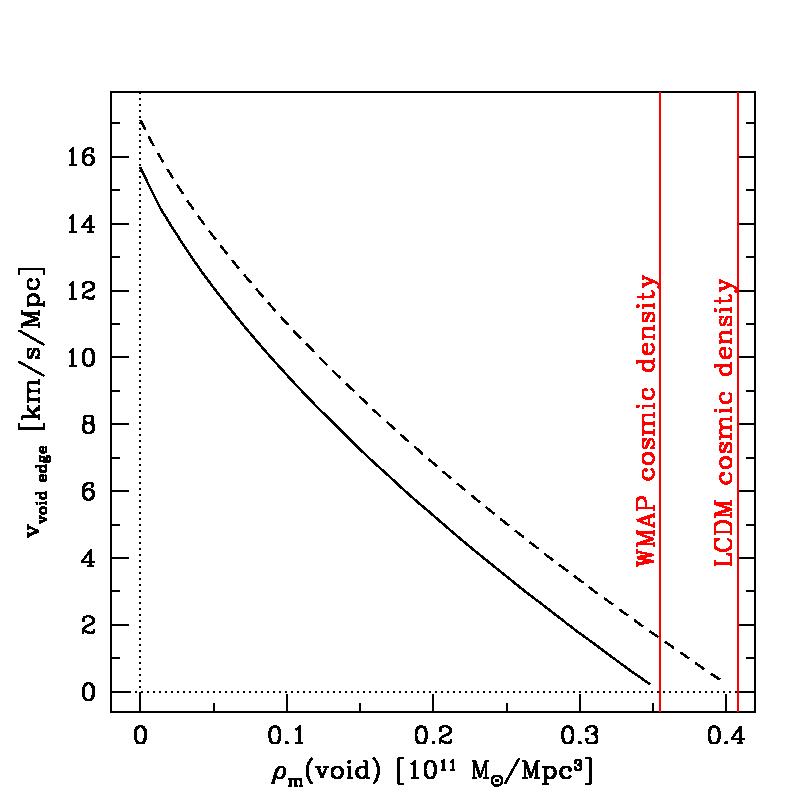}
\caption{
Velocities of expansion generated in each Mpc in a void as a function of the density in the void.
The solid and dashed lines corresponds to the cosmologies with H$_0=73$, $\Omega_m=0.24$ (WMAP) and H$_0=70$, $\Omega_m=0.3$ (LCDM), respectively.    
}
\label{void_veloc}
\end{center}
\end{figure}

These values are in good agreement with results from simulations reported by \citet{2007arXiv0708.1441V}.  Those simulations show that in models with $\Lambda = 0$ the voids are not fully evacuated at the present epoch and motions out of voids are consequently lower than if the voids were empty.  However in the simulations with $\Omega_{\Lambda} \sim 0.7$ the voids are quite empty at $z=0$, suggesting that we can give serious consideration to this possibility in the case of the Local Void. Figure~\ref{void_veloc} illustrates the dependence of outflow velocities on the residual density within a void for two cosmological models. 

In Section~3.4 it was determined that the Local Sheet has a bulk motion of 259~\kms\ away from the Local Void.  Simplistically, it could be inferred that  the radius of a completely empty Local Void is at least 16 Mpc.

It is not out of the question that the entire Local Void including Inner, North, and South components could have this dimension.  The geometry of the Local Void is poorly delineated because of the unfortunate intersection of the plane of the Milky Way.  Certainly, this region is not entirely empty.  Several wispy filaments lace through the volume.  If the void is not empty then a larger size is required to generate the observed expansion velocities.

We can raise a couple of layers of complexity.  Recovery of velocity fields from simulations provide a guide.  Figure~\ref{dtfe} is extracted from \citet{2007arXiv0708.1441V}.  The example we show involves the reasonably symmetric convergence of material onto a filament.  Streaming motions grow approximately linearly away from the centers of the low density regions on each side of the filament, reaching maxima at the interface with the filament.  This behavior is consistent with expectations based on the formulae given above.  However, although particles impinging on the filament have large velocities in the example seen here, the filament does not have a large lateral bulk motion.  In this particular case, there is considerable symmetry with the influx from the opposite sides of the filament.  In other circumstances, there might be an asymmetry.   An example is provided by Figure~\ref{horizon}.  The sequence of 4 time steps is drawn from an N-body GADGET simulation \citep{2005MNRAS.364.1105S}   produced by the HORIZON collaboration (www.projet-horizon.fr).  Attention is drawn to a filament that has the combined properties seen in the Local Sheet of lateral
motion due to the dominance of a void on one side and a flow toward a nearby cluster in an orthogonal direction.  Two adjacent filaments displace laterally toward convergence, reproducing the behavior seen between the Local Sheet and the Leo Spur.

\begin{figure}[htbp]
\figurenum{17}
\begin{center}
\plotone{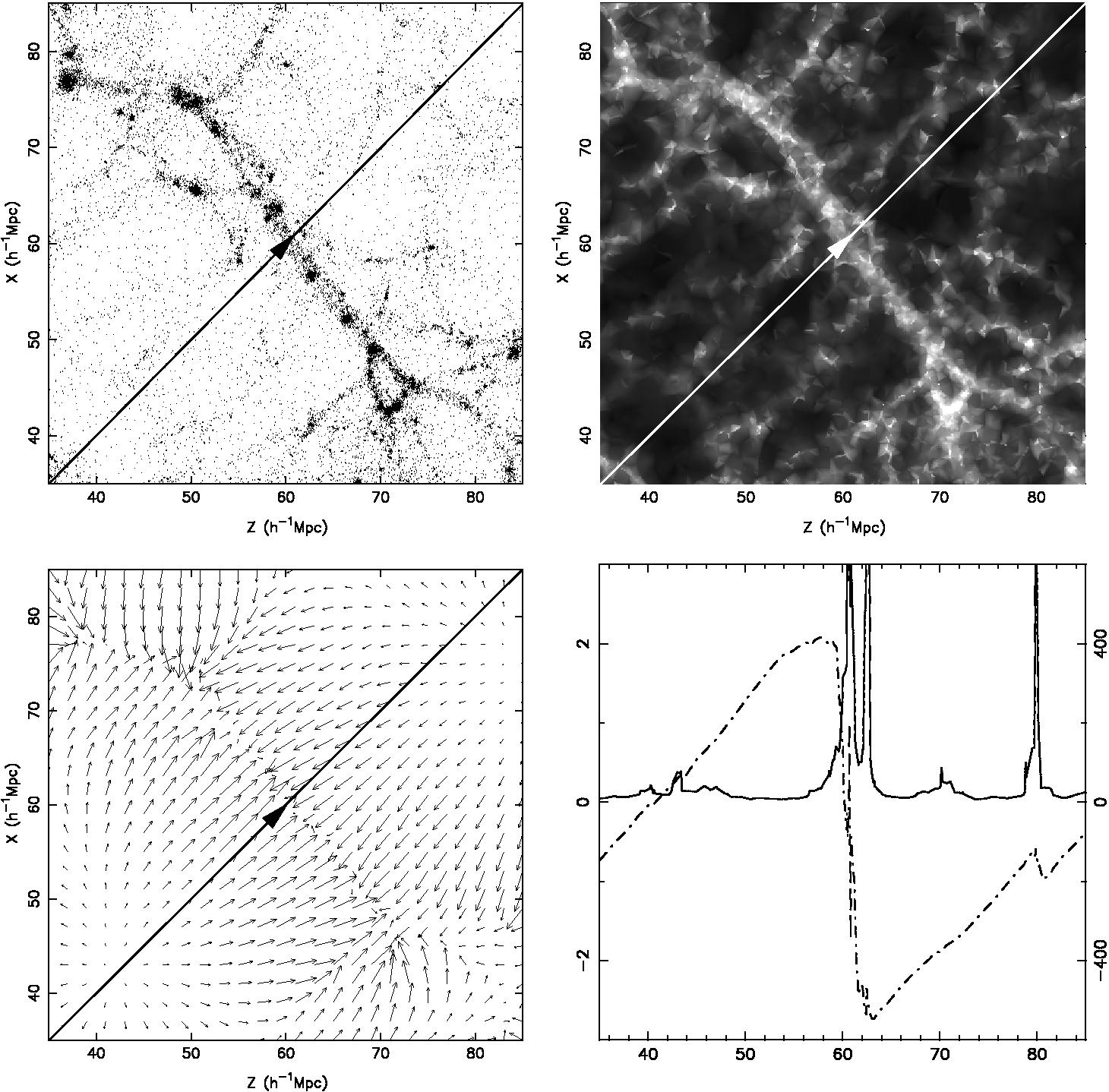}
\caption{Example of void dynamics extracted from thesis by Schaap.  Top left: particle distribution in a thin slice through simulation box.  Top right: 2D slice through the 3D Delaunay Tessellation Field Estimator (DTFE) density field reconstruction of the simulation.  Bottom left: 2D slice through the 3D DTFE velocity field reconstruction.  Bottom right: reconstructions along the thick line shown in the other frames; solid line is the density reconstruction and dot--dashed line is the velocity reconstruction.  Velocities reach large amplitudes toward the edges of the voids. 
}
\label{dtfe}
\end{center}
\end{figure}

\begin{figure}[htbp]
\figurenum{18}
\begin{center}
\plotone{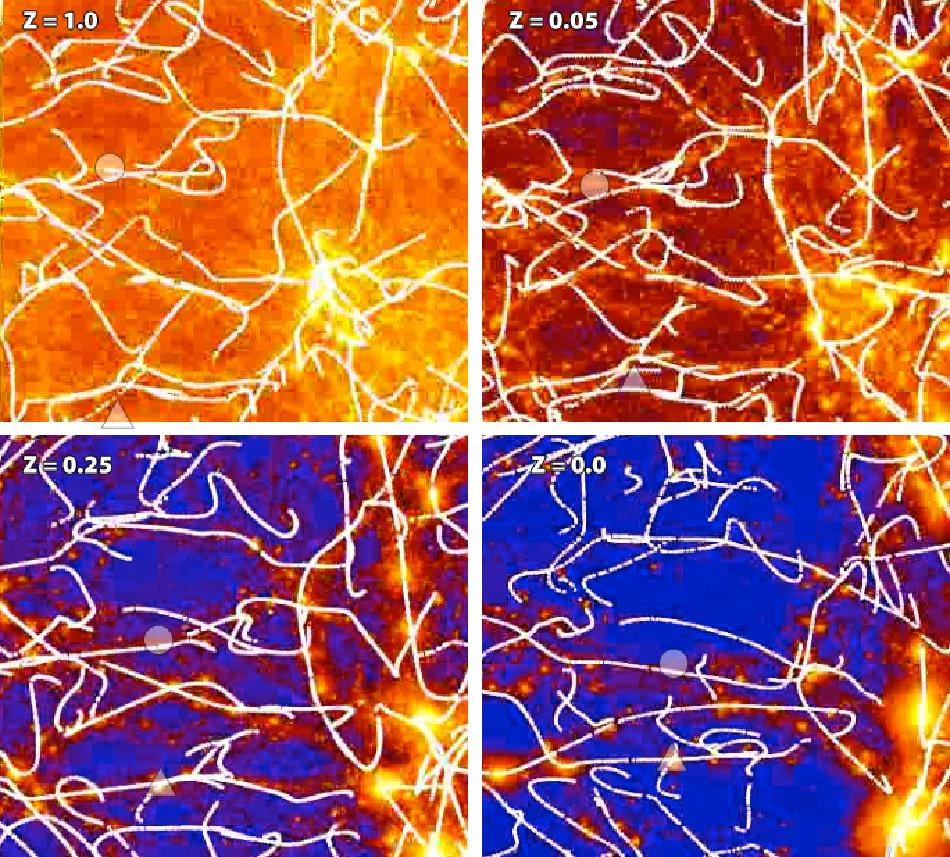}
\caption{Four snapshots in time of an N-body simulation with conditions resembling the observations.  The dark matter GADGET simulation is of a 20 Mpc box with $256^3$ particles, $\Lambda$CDM, with $\Omega_m=0.3$ and $\Omega_{\Lambda}=0.7$.  The region shown is 10 Mpc across in co-moving coordinates at the redshift steps $z=1$, 0.5, 0.25, and 0.  The skeleton method  \citep{2006astro.ph..2628S} delineates the backbone of the filaments at each step.  The large circle in each panel identifies the progression of a location that ends up with properties resembling those of the Milky Way:  on a filament, with motions that are both lateral to the filament away from a void, and along the filament toward a cluster.   The large triangle in each panel tracks a location that comes to resemble the Leo Spur, with upward motions headed toward a future closure with the filament bearing the circle. For animation, see
ifa.hawaii.edu/$\sim$tully/skeleton-060421.mpg.
}
\label{horizon}
\end{center}
\end{figure}

The observed motion of ESO~461--36 gives a useful constraint.  This dwarf galaxy is still well within the Local Void, still within the unfettered flow toward our filament.  With reference to the flow pattern seen in the lower right panel of Fig.~\ref{dtfe}, a galaxy such as ESO~461-36 might be anticipated to be near the maximum of the swing of peculiar velocities.  We measure a peculiar velocity for this galaxy of $-30$~\kms.  As was noted in Section 3.5, this motion is additive with our velocity in the reference frame of the Local Supercluster, which implies a peculiar velocity of at least 350~\kms\ with respect to that reference frame.  The peculiar velocity is higher if the galaxy has a significant tangential component.  This galaxy should have a motion of $\sim 120$~\kms\ due to the influence of the Virgo Cluster, leaving an additional $\sim 230$~\kms\  attributable to evacuation from the void.  Equation (30) requires 
that ESO~461--36 be at least 17 Mpc from the void center.  Since we are 6 Mpc farther back, that would put us 23 Mpc from the void center.

Maybe so.  However, this requires that the void be very big and very empty.  The situation invites consideration of more radical alternatives.   \citet{2007PhRvD..75f3507D}  argue that dark energy with a varying equation of state might have enhanced density in places with lower mass density; i.e., in voids.  A consequence could be an enhanced expansion rate of space in voids resulting in increased velocities away from void centers.  Potentially, observations of the motions of galaxies within voids could give important information about the fundamental properties of the universe.

\section{Summary}

Our motion inferred by the CMB dipole anisotropy of 631~\kms\ in the Local Sheet reference frame decomposes into three main contributions that are almost orthogonal.  The three main components conveniently lie near the cardinal axes of the Supergalactic coordinate system (only partially by chance). Two of these components decouple from the third because two are local, the closest seen abruptly in a peculiar velocity discontinuity at 7~Mpc, while the third is large scale, acquiring an importance at $V_{LS} > 3000$~\kms.

One of the local components is caused by the Virgo Cluster and its dense surroundings.  The motion of the Local Sheet with respect to galaxies within 3000~\kms\ has a component of 185~\kms\ toward this cluster at $L=103, B=-2$.  The residual from the local component of CMB motion is a Local Sheet velocity of 259~\kms\ toward $L=11, B=-72$, toward nothing of importance but {\it away} from the Local Void.  Subtraction of these two local components from the CMB vector leaves the third component attributed to large scale attractors of 455~\kms\ toward $L=162, B=-16$, close to the direction of the Centaurus Cluster.  These three components cause motions roughly aligned with the +SGY, $-$SGZ, and $-$SGX axes respectively, providing a decomposition that is gruntling.

The availability of a large number of accurate TRGB distances has clarified details about the `local velocity anomaly'.  Our Local Sheet is participating in the cosmic expansion (though probably somewhat retarded) but simultaneously moving in bulk toward the Virgo Cluster and away from the Local Void.  Our Local Group has only a small peculiar velocity (66~\kms) with respect to other galaxies of the Local Sheet which, internally, has only small random motions (40~\kms\ in the radial direction averaged over groups).   We advocate the use of the Local Sheet as a frame of reference in preference to the Local Group because the reference sample is five times larger and more widely distributed.  The local velocity anomaly is given emphasis because there is a sharp discontinuity in velocities as we look beyond the Local Sheet  toward $-$SGZ.  Galaxies in the Leo Spur with well measured distances all have large negative peculiar velocities.  Much, if not all, of these motions are a reflex of the motion of the Local Sheet toward $-$SGZ and away from the Local Void.  

Our distances are not yet numerous or accurate enough to demonstrate if other filaments have similar bulk motions.  However we do see clearly that a galaxy within but near our edge of the Local Void, ESO~461$-$36, has a peculiar velocity of at least 230~\kms\ away from the center of the void.  Our  Local Sheet and that galaxy are participating in the evacuation of the Local Void.
The large expulsion velocities imply a dimension of the radius to the center of the Local Void  at our position of  at least 23~Mpc.  We lie on the boundary of a major void.  The evidence for expansion is unambiguous at our privileged position on the void wall.  Voids have few galaxies but are they really empty? Yes, to create such a large outflow, our Local Void must be really empty.

\vskip 1cm
\noindent
Figure~\ref{dtfe} is originally from the Ph.D. thesis by W.E. Schaap and was provided by Rien van de Weygaert.
This research has been sponsored by Space Telescope Science Institute awards GO-9771, GO-9950, GO-10210, GO-10905, and the National Science Foundation award AST 03-07706.  IK is partially supported by the programs DFG-RFBR 06-02-04017 and RFBR 07-02-00005 in Russia.

\bibliography{paper}




\bibliographystyle{apj}

\clearpage

{\scriptsize
\begin{table}
\tablenum{3}
\caption{Reference Frame Transformations}
\begin{tabular}{lrlrlrllll}

\label{tbl:vel_trans}
 & & & & & & &  &  & \\
\hline
\hline
\multicolumn{10}{l}{Transform velocities w.r.t. Sun to velocities w.r.t. Galactic center: $V_{GSR}-V_{\odot}$}\\
$V_{GSR}^{\odot}$ & $+9.3$ & $\hat x$ & +218 & $\hat y$ & $+7.6$ & $\hat z$ &  $=219 \pm 12$ & $\ell=87.6 \pm 1$ & $b=+2 \pm 1$ \\

                                    & $+140$ & $\hat X$ & $-10$ &  $\hat Y$ & +167  & $\hat Z$ &                           & $L=356$         & $B=+50$      \\
\hline
\multicolumn{10}{l}{Transform velocities w.r.t. Sun to velocities w.r.t. Local Group: $V_{LG}-V_{\odot}$}\\
$V_{LG}^{\odot}$ & $-86$ & $\hat x$ & +305 & $\hat y$ & $-33$ & $\hat z$ &  $=318 \pm 20$ & $\ell=106 \pm 4$ & $b=-6 \pm 4$ \\

                                & $+270$ & $\hat X$ & $-52$ &  $\hat Y$ & +159  & $\hat Z$ &                        & $L=349$         & $B=+30$      \\
\hline
\multicolumn{10}{l}{Transform velocities w.r.t. Sun to velocities w.r.t. Local Sheet: $V_{LS}-V_{\odot}$}\\
$V_{LS}^{\odot}$ & $-26$ & $\hat x$ & +317 & $\hat y$ & $-8$ & $\hat z$ &  $=318 \pm 20$ & $\ell=95 \pm 4$ & $b=-1 \pm 4$ \\

                                & $+234$ & $\hat X$ & $-31$ &  $\hat Y$ & +214  & $\hat Z$ &                   & $L=353$         & $B=+42$      \\
\hline
\multicolumn{10}{l}{Motion of Galactic center in Local Group rest frame: $V_{LG}^{GSR}=V_{LG}-V_{GSR}$}\\
$V_{LG}^{GSR} $ & $-95$ & $\hat x$ & +87 & $\hat y$ & $-41$ & $\hat z$ &  $=135 \pm 25$ & $\ell=137 \pm 10$ & $b=-18 \pm 10$ \\

                                 & $+129$ & $\hat X$ & $-42$ &  $\hat Y$ & -8  & $\hat Z$ &                & $L=342$           & $B=-3$         \\
\hline
\multicolumn{10}{l}{Motion of the Local Group w.r.t. the Local Sheet: $V_{LS}-V_{LG}$}\\
$V_{LS}^{LG} $ & $+60$ & $\hat x$ & +12 & $\hat y$ & $+25$ & $\hat z$ &  $=66 \pm 24$  & $\ell=349$ & $b=+22$ \\

                              & $-34$ & $\hat X$ & +20 & $\hat Y$ & +53 & $\hat Z$ &         & $L=150 \pm 37$ & $B=+53 \pm 20$ \\
\hline
\multicolumn{10}{l}{Transform velocities w.r.t. Local Sheet to velocities w.r.t. galaxies with $V_{LS}<3000$~km s$^{-1}$}\\
$V_{LSC}^{LS}$ & $-211$ & $\hat x$ & $-178$ & $\hat y$ & $+169$ & $\hat z$ &  $=323 \pm 25 $ & $\ell=220 \pm 7$ & $b=+32 \pm 6$ \\

                               & $+35  $ & $\hat X$      & $+196$ &  $\hat Y$    & $-255$ & $\hat Z$ &               & $L=80$           & $B=-52$       \\
\hline
\multicolumn{10}{l}{Component of $V_{LSC}^{LS}$ directed toward Virgo Cluster}\\
$V_{LSC;V}^{LS}$ & $+11$ & $\hat x$ & $-48$ & $\hat y$ & $+179$ & $\hat z$ &  $=185 \pm 20 $ & $\ell=283.8$ & $b=+74.5$ \\

                          & $-42 $ & $\hat X$      & $+180$ &  $\hat Y$    & $-7$ & $\hat Z$ &             & $L=102.9$           & $B=-2.4$       \\
\hline
\multicolumn{10}{l}{Motion of the Local Sheet away from  the Local Void, $V_{LSC;LV}^{LS}=V_{LSC}^{LS}-V_{LSC;V}^{LS}$}\\
$V_{LSC;LV}^{LS}$ & $-222$ & $\hat x$ & $-130$ & $\hat y$ & $-10$   & $\hat z$ &  $=259 \pm 25 $ & $\ell=210 \pm 7$ & $b=-2 \pm 6$ \\

                               & $+77$ & $\hat X$      & $+16$ &  $\hat Y$    & $-248$   & $\hat Z$ &          & $L=11$      & $B=-72$ \\
\hline
\multicolumn{10}{l}{Transform velocities w.r.t. Local Sheet to velocities w.r.t CMB reference frame}\\
$V_{CMB}^{LS}$ & $+1$ & $\hat x$ & $-563$ & $\hat y$ & $+285$   & $\hat z$ &  $=631 \pm 20 $ & $\ell=270 \pm 3$ & $b=+27 \pm 3$ \\

                                & $-381$ & $\hat X$      & $+331$ &  $\hat Y$    & $-380$   & $\hat Z$ &         & $L=139$      & $B=-37$ \\
\hline
\multicolumn{10}{l}{Large scale component of motion w.r.t the Local Sheet: $V_{CMB;LSS}^{LS}=V_{CMB}^{LS}-V_{LSC}^{LS}$}\\
$V_{CMB;LSS}^{LS}$ & $+212$ & $\hat x$ & $-385$ & $\hat y$ & $+116$   & $\hat z$ &  $=455 \pm 15 $ & $\ell=299 \pm 3$ & $b=+15 \pm 3$ \\

                               & $-416$ & $\hat X$      & $+135$ &  $\hat Y$    & $-125$   & $\hat Z$ &           & $L=162$      & $B=-16$ \\
\hline
\hline
\end{tabular}
Glossary of super/sub-script acronyms:

\scriptsize{
 $\odot$~~~~~Sun
 
CMB~Cosmic Microwave Background

GSR~Galactic Standard of Rest 

LG~~~Local Group

LV~~~Local Void

LS~~~Local Sheet

LSC~Local Supercluster ($<3000$~\kms)

LSS~Large Scale Structure ($>3000$~\kms)

V~~~~Virgo Cluster
}
\end{table}
}

\end{document}

%% file: defs.tex
\newcommand{\kms}{km~s$^{-1}$}
\newcommand{\Msun}{M_{\odot}}
\newcommand{\Lsun}{L_{\odot}}
\newcommand{\ML}{M_{\odot}/L_{\odot}}
\newcommand{\etal}{{et al.}\ }
\newcommand{\hhh}{h_{100}}
\newcommand{\hsq}{h_{100}^{-2}}
\newcommand{\tn}{\tablenotemark}
\newcommand{\mdot}{\dot{M}}
\newcommand{\p}{^\prime}
\newcommand{\kmsMpc}{km~s$^{-1}$~Mpc$^{-1}$}

%% file: ms.bbl
\begin{thebibliography}{58}
\expandafter\ifx\csname natexlab\endcsname\relax\def\natexlab#1{#1}\fi

\bibitem[{{Aaronson} {et~al.}(1986){Aaronson}, {Bothun}, {Mould}, {Huchra},
  {Schommer}, \& {Cornell}}]{1986ApJ...302..536A}
{Aaronson}, M., {Bothun}, G., {Mould}, J., {Huchra}, J., {Schommer}, R.~A., \&
  {Cornell}, M.~E. 1986, \apj, 302, 536

\bibitem[{{Aaronson} {et~al.}(1982){Aaronson}, {Huchra}, {Mould}, {Schechter},
  \& {Tully}}]{1982ApJ...258...64A}
{Aaronson}, M., {Huchra}, J., {Mould}, J., {Schechter}, P.~L., \& {Tully},
  R.~B. 1982, \apj, 258, 64

\bibitem[{{Bouchet} \& {Lachieze-Rey}(1993)}]{1993cvf..conf.....B}
{Bouchet}, F. \& {Lachieze-Rey}, M., eds. 1993, {Cosmic velocity fields}

\bibitem[{{Courteau} {et~al.}(2000){Courteau}, {Strauss}, \&
  {Willick}}]{2000ASPC..201.....C}
{Courteau}, S., {Strauss}, M.~A., \& {Willick}, J., eds. 2000, {Cosmic Flows
  Workshop}

\bibitem[{{Courteau} \& {van den Bergh}(1999)}]{1999AJ....118..337C}
{Courteau}, S. \& {van den Bergh}, S. 1999, \aj, 118, 337

\bibitem[{{de Vaucouleurs} {et~al.}(1991){de Vaucouleurs}, {de Vaucouleurs},
  {Corwin}, {Buta}, {Paturel}, \& {Fouqu{\'e}}}]{1991trcb.book.....D}
{de Vaucouleurs}, G., {de Vaucouleurs}, A., {Corwin}, Jr., H.~G., {Buta},
  R.~J., {Paturel}, G., \& {Fouqu{\'e}}, P. 1991, {Third Reference Catalogue of
  Bright Galaxies} (Volume 1-3, XII, 2069 pp.~7 figs..~ Springer-Verlag Berlin
  Heidelberg New York)

\bibitem[{{Dolphin} {et~al.}(2003){Dolphin}, {Saha}, {Skillman}, {Dohm-Palmer},
  {Tolstoy}, {Cole}, {Gallagher}, {Hoessel}, \& {Mateo}}]{2003AJ....125.1261D}
{Dolphin}, A.~E., {Saha}, A., {Skillman}, E.~D., {Dohm-Palmer}, R.~C.,
  {Tolstoy}, E., {Cole}, A.~A., {Gallagher}, J.~S., {Hoessel}, J.~G., \&
  {Mateo}, M. 2003, \aj, 125, 1261

\bibitem[{{Dutta} \& {Maor}(2007)}]{2007PhRvD..75f3507D}
{Dutta}, S. \& {Maor}, I. 2007, \prd, 75, 063507

\bibitem[{{Eisenhauer} {et~al.}(2005){Eisenhauer}, {Genzel}, {Alexander},
  {Abuter}, {Paumard}, {Ott}, {Gilbert}, {Gillessen}, {Horrobin}, {Trippe},
  {Bonnet}, {Dumas}, {Hubin}, {Kaufer}, {Kissler-Patig}, {Monnet},
  {Str{\"o}bele}, {Szeifert}, {Eckart}, {Sch{\"o}del}, \&
  {Zucker}}]{2005ApJ...628..246E}
{Eisenhauer}, F., {Genzel}, R., {Alexander}, T., {Abuter}, R., {Paumard}, T.,
  {Ott}, T., {Gilbert}, A., {Gillessen}, S., {Horrobin}, M., {Trippe}, S.,
  {Bonnet}, H., {Dumas}, C., {Hubin}, N., {Kaufer}, A., {Kissler-Patig}, M.,
  {Monnet}, G., {Str{\"o}bele}, S., {Szeifert}, T., {Eckart}, A.,
  {Sch{\"o}del}, R., \& {Zucker}, S. 2005, \apj, 628, 246

\bibitem[{{Erdo{\u g}du} {et~al.}(2006{\natexlab{a}}){Erdo{\u g}du}, {Huchra},
  {Lahav}, {Colless}, {Cutri}, {Falco}, {George}, {Jarrett}, {Jones},
  {Kochanek}, {Macri}, {Mader}, {Martimbeau}, {Pahre}, {Parker}, {Rassat}, \&
  {Saunders}}]{2006MNRAS.368.1515E}
{Erdo{\u g}du}, P., {Huchra}, J.~P., {Lahav}, O., {Colless}, M., {Cutri},
  R.~M., {Falco}, E., {George}, T., {Jarrett}, T., {Jones}, D.~H., {Kochanek},
  C.~S., {Macri}, L., {Mader}, J., {Martimbeau}, N., {Pahre}, M., {Parker}, Q.,
  {Rassat}, A., \& {Saunders}, W. 2006{\natexlab{a}}, \mnras, 368, 1515

\bibitem[{{Erdo{\u g}du} {et~al.}(2006{\natexlab{b}}){Erdo{\u g}du}, {Lahav},
  {Huchra}, {Colless}, {Cutri}, {Falco}, {George}, {Jarrett}, {Jones}, {Macri},
  {Mader}, {Martimbeau}, {Pahre}, {Parker}, {Rassat}, \&
  {Saunders}}]{2006MNRAS.373...45E}
{Erdo{\u g}du}, P., {Lahav}, O., {Huchra}, J.~P., {Colless}, M., {Cutri},
  R.~M., {Falco}, E., {George}, T., {Jarrett}, T., {Jones}, D.~H., {Macri},
  L.~M., {Mader}, J., {Martimbeau}, N., {Pahre}, M.~A., {Parker}, Q.~A.,
  {Rassat}, A., \& {Saunders}, W. 2006{\natexlab{b}}, \mnras, 373, 45

\bibitem[{{Evans} \& {Wilkinson}(2000)}]{2000MNRAS.316..929E}
{Evans}, N.~W. \& {Wilkinson}, M.~I. 2000, \mnras, 316, 929

\bibitem[{{Faber} \& {Burstein}(1988)}]{1988lsmu.book..115F}
{Faber}, S.~M. \& {Burstein}, D. 1988, {Motions of galaxies in the neighborhood
  of the local group} (Large-Scale Motions in the Universe: A Vatican study
  Week), 115--167

\bibitem[{{Feast} \& {Whitelock}(1997)}]{1997MNRAS.291..683F}
{Feast}, M. \& {Whitelock}, P. 1997, \mnras, 291, 683

\bibitem[{{Fixsen} {et~al.}(1996){Fixsen}, {Cheng}, {Gales}, {Mather},
  {Shafer}, \& {Wright}}]{1996ApJ...473..576F}
{Fixsen}, D.~J., {Cheng}, E.~S., {Gales}, J.~M., {Mather}, J.~C., {Shafer},
  R.~A., \& {Wright}, E.~L. 1996, \apj, 473, 576

\bibitem[{{Freedman} {et~al.}(2001){Freedman}, {Madore}, {Gibson}, {Ferrarese},
  {Kelson}, {Sakai}, {Mould}, {Kennicutt}, {Ford}, {Graham}, {Huchra},
  {Hughes}, {Illingworth}, {Macri}, \& {Stetson}}]{2001ApJ...553...47F}
{Freedman}, W.~L., {Madore}, B.~F., {Gibson}, B.~K., {Ferrarese}, L., {Kelson},
  D.~D., {Sakai}, S., {Mould}, J.~R., {Kennicutt}, Jr., R.~C., {Ford}, H.~C.,
  {Graham}, J.~A., {Huchra}, J.~P., {Hughes}, S.~M.~G., {Illingworth}, G.~D.,
  {Macri}, L.~M., \& {Stetson}, P.~B. 2001, \apj, 553, 47

\bibitem[{{Hoffman} \& {Salpeter}(1982)}]{1982ApJ...263..485H}
{Hoffman}, G.~L. \& {Salpeter}, E.~E. 1982, \apj, 263, 485

\bibitem[{{Iwata} {et~al.}(2005){Iwata}, {Ohta}, {Nakanishi}, {Chamaraux}, \&
  {Roman}}]{2005ASPC..329...59I}
{Iwata}, I., {Ohta}, K., {Nakanishi}, K., {Chamaraux}, P., \& {Roman}, A.~T.
  2005, in Astronomical Society of the Pacific Conference Series, Vol. 329,
  Nearby Large-Scale Structures and the Zone of Avoidance, ed. A.~P. {Fairall}
  \& P.~A. {Woudt}, 59--+

\bibitem[{{Karachentsev} {et~al.}(2006){Karachentsev}, {Dolphin}, {Tully},
  {Sharina}, {Makarova}, {Makarov}, {Karachentseva}, {Sakai}, \&
  {Shaya}}]{2006AJ....131.1361K}
{Karachentsev}, I.~D., {Dolphin}, A., {Tully}, R.~B., {Sharina}, M.,
  {Makarova}, L., {Makarov}, D., {Karachentseva}, V., {Sakai}, S., \& {Shaya},
  E.~J. 2006, \aj, 131, 1361

\bibitem[{{Karachentsev} {et~al.}(2004){Karachentsev}, {Karachentseva},
  {Huchtmeier}, \& {Makarov}}]{2004AJ....127.2031K}
{Karachentsev}, I.~D., {Karachentseva}, V.~E., {Huchtmeier}, W.~K., \&
  {Makarov}, D.~I. 2004, \aj, 127, 2031

\bibitem[{{Karachentsev} \& {Makarov}(1996)}]{1996AJ....111..794K}
{Karachentsev}, I.~D. \& {Makarov}, D.~A. 1996, \aj, 111, 794

\bibitem[{{Karachentsev} {et~al.}(2003){Karachentsev}, {Makarov}, {Sharina},
  {Dolphin}, {Grebel}, {Geisler}, {Guhathakurta}, {Hodge}, {Karachentseva},
  {Sarajedini}, \& {Seitzer}}]{2003A&A...398..479K}
{Karachentsev}, I.~D., {Makarov}, D.~I., {Sharina}, M.~E., {Dolphin}, A.~E.,
  {Grebel}, E.~K., {Geisler}, D., {Guhathakurta}, P., {Hodge}, P.~W.,
  {Karachentseva}, V.~E., {Sarajedini}, A., \& {Seitzer}, P. 2003, \aap, 398,
  479

\bibitem[{{Karachentsev} {et~al.}(2002{\natexlab{a}}){Karachentsev},
  {Mitronova}, {Karachentseva}, {Kudrya}, \& {Jarrett}}]{2002A&A...396..431K}
{Karachentsev}, I.~D., {Mitronova}, S.~N., {Karachentseva}, V.~E., {Kudrya},
  Y.~N., \& {Jarrett}, T.~H. 2002{\natexlab{a}}, \aap, 396, 431

\bibitem[{{Karachentsev} {et~al.}(2002{\natexlab{b}}){Karachentsev}, {Sharina},
  {Makarov}, {Dolphin}, {Grebel}, {Geisler}, {Guhathakurta}, {Hodge},
  {Karachentseva}, {Sarajedini}, \& {Seitzer}}]{2002A&A...389..812K}
{Karachentsev}, I.~D., {Sharina}, M.~E., {Makarov}, D.~I., {Dolphin}, A.~E.,
  {Grebel}, E.~K., {Geisler}, D., {Guhathakurta}, P., {Hodge}, P.~W.,
  {Karachentseva}, V.~E., {Sarajedini}, A., \& {Seitzer}, P.
  2002{\natexlab{b}}, \aap, 389, 812

\bibitem[{{Kocevski} \& {Ebeling}(2006)}]{2006ApJ...645.1043K}
{Kocevski}, D.~D. \& {Ebeling}, H. 2006, \apj, 645, 1043

\bibitem[{{Lahav} {et~al.}(1988){Lahav}, {Lynden-Bell}, \&
  {Rowan-Robinson}}]{1988MNRAS.234..677L}
{Lahav}, O., {Lynden-Bell}, D., \& {Rowan-Robinson}, M. 1988, \mnras, 234, 677

\bibitem[{{Lynden-Bell} {et~al.}(1988){Lynden-Bell}, {Faber}, {Burstein},
  {Davies}, {Dressler}, {Terlevich}, \& {Wegner}}]{1988ApJ...326...19L}
{Lynden-Bell}, D., {Faber}, S.~M., {Burstein}, D., {Davies}, R.~L., {Dressler},
  A., {Terlevich}, R.~J., \& {Wegner}, G. 1988, \apj, 326, 19

\bibitem[{{Makarov} {et~al.}(2006){Makarov}, {Makarova}, {Rizzi}, {Tully},
  {Dolphin}, {Sakai}, \& {Shaya}}]{2006AJ....132.2729M}
{Makarov}, D., {Makarova}, L., {Rizzi}, L., {Tully}, R.~B., {Dolphin}, A.~E.,
  {Sakai}, S., \& {Shaya}, E.~J. 2006, \aj, 132, 2729

\bibitem[{{Maller} {et~al.}(2003){Maller}, {McIntosh}, {Katz}, \&
  {Weinberg}}]{2003ApJ...598L...1M}
{Maller}, A.~H., {McIntosh}, D.~H., {Katz}, N., \& {Weinberg}, M.~D. 2003,
  \apjl, 598, L1

\bibitem[{{Mei} {et~al.}(2007){Mei}, {Blakeslee}, {C{\^o}t{\'e}}, {Tonry},
  {West}, {Ferrarese}, {Jord{\'a}n}, {Peng}, {Anthony}, \&
  {Merritt}}]{2007ApJ...655..144M}
{Mei}, S., {Blakeslee}, J.~P., {C{\^o}t{\'e}}, P., {Tonry}, J.~L., {West},
  M.~J., {Ferrarese}, L., {Jord{\'a}n}, A., {Peng}, E.~W., {Anthony}, A., \&
  {Merritt}, D. 2007, \apj, 655, 144

\bibitem[{{Meyer} {et~al.}(2004){Meyer}, {Zwaan}, {Webster}, {Staveley-Smith},
  {Ryan-Weber}, {Drinkwater}, {Barnes}, {Howlett}, {Kilborn}, {Stevens},
  {Waugh}, {Pierce}, {Bhathal}, {de Blok}, {Disney}, {Ekers}, {Freeman},
  {Garcia}, {Gibson}, {Harnett}, {Henning}, {Jerjen}, {Kesteven}, {Knezek},
  {Koribalski}, {Mader}, {Marquarding}, {Minchin}, {O'Brien}, {Oosterloo},
  {Price}, {Putman}, {Ryder}, {Sadler}, {Stewart}, {Stootman}, \&
  {Wright}}]{2004MNRAS.350.1195M}
{Meyer}, M.~J., {Zwaan}, M.~A., {Webster}, R.~L., {Staveley-Smith}, L.,
  {Ryan-Weber}, E., {Drinkwater}, M.~J., {Barnes}, D.~G., {Howlett}, M.,
  {Kilborn}, V.~A., {Stevens}, J., {Waugh}, M., {Pierce}, M.~J., {Bhathal}, R.,
  {de Blok}, W.~J.~G., {Disney}, M.~J., {Ekers}, R.~D., {Freeman}, K.~C.,
  {Garcia}, D.~A., {Gibson}, B.~K., {Harnett}, J., {Henning}, P.~A., {Jerjen},
  H., {Kesteven}, M.~J., {Knezek}, P.~M., {Koribalski}, B.~S., {Mader}, S.,
  {Marquarding}, M., {Minchin}, R.~F., {O'Brien}, J., {Oosterloo}, T., {Price},
  R.~M., {Putman}, M.~E., {Ryder}, S.~D., {Sadler}, E.~M., {Stewart}, I.~M.,
  {Stootman}, F., \& {Wright}, A.~E. 2004, \mnras, 350, 1195

\bibitem[{{Mohayaee} \& {Tully}(2005)}]{2005ApJ...635L.113M}
{Mohayaee}, R. \& {Tully}, R.~B. 2005, \apjl, 635, L113

\bibitem[{{Raychaudhury}(1989)}]{1989Natur.342..251R}
{Raychaudhury}, S. 1989, \nat, 342, 251

\bibitem[{{Rizzi} {et~al.}(2007){Rizzi}, {Tully}, {Makarov}, {Makarova},
  {Dolphin}, {Sakai}, \& {Shaya}}]{2007ApJ...661..815R}
{Rizzi}, L., {Tully}, R.~B., {Makarov}, D., {Makarova}, L., {Dolphin}, A.~E.,
  {Sakai}, S., \& {Shaya}, E.~J. 2007, \apj, 661, 815

\bibitem[{{Rubin} \& {Coyne}(1988)}]{1988lsmu.book.....R}
{Rubin}, V.~C. \& {Coyne}, G.~V. 1988, {Large-scale motions in the universe}
  (Large-Scale Motions in the Universe: A Vatican study Week)

\bibitem[{{Sakai} {et~al.}(2004){Sakai}, {Ferrarese}, {Kennicutt}, \&
  {Saha}}]{2004ApJ...608...42S}
{Sakai}, S., {Ferrarese}, L., {Kennicutt}, Jr., R.~C., \& {Saha}, A. 2004,
  \apj, 608, 42

\bibitem[{{Sakai} {et~al.}(1996){Sakai}, {Madore}, \&
  {Freedman}}]{1996ApJ...461..713S}
{Sakai}, S., {Madore}, B.~F., \& {Freedman}, W.~L. 1996, \apj, 461, 713

\bibitem[{{Scaramella} {et~al.}(1989){Scaramella}, {Baiesi-Pillastrini},
  {Chincarini}, {Vettolani}, \& {Zamorani}}]{1989Natur.338..562S}
{Scaramella}, R., {Baiesi-Pillastrini}, G., {Chincarini}, G., {Vettolani}, G.,
  \& {Zamorani}, G. 1989, \nat, 338, 562

\bibitem[{{Schlegel} {et~al.}(1998){Schlegel}, {Finkbeiner}, \&
  {Davis}}]{1998ApJ...500..525S}
{Schlegel}, D.~J., {Finkbeiner}, D.~P., \& {Davis}, M. 1998, \apj, 500, 525

\bibitem[{{Shaya}(1984)}]{1984ApJ...280..470S}
{Shaya}, E.~J. 1984, \apj, 280, 470

\bibitem[{{Shaya} {et~al.}(1995){Shaya}, {Peebles}, \&
  {Tully}}]{1995ApJ...454...15S}
{Shaya}, E.~J., {Peebles}, P.~J.~E., \& {Tully}, R.~B. 1995, \apj, 454, 15

\bibitem[{{Sousbie} {et~al.}(2006){Sousbie}, {Pichon}, {Courtois}, {Colombi},
  \& {Novikov}}]{2006astro.ph..2628S}
{Sousbie}, T., {Pichon}, C., {Courtois}, H., {Colombi}, S., \& {Novikov}, D.
  2006, ArXiv Astrophysics e-prints

\bibitem[{{Spergel} {et~al.}(2003){Spergel}, {Verde}, {Peiris}, {Komatsu},
  {Nolta}, {Bennett}, {Halpern}, {Hinshaw}, {Jarosik}, {Kogut}, {Limon},
  {Meyer}, {Page}, {Tucker}, {Weiland}, {Wollack}, \&
  {Wright}}]{2003ApJS..148..175S}
{Spergel}, D.~N., {Verde}, L., {Peiris}, H.~V., {Komatsu}, E., {Nolta}, M.~R.,
  {Bennett}, C.~L., {Halpern}, M., {Hinshaw}, G., {Jarosik}, N., {Kogut}, A.,
  {Limon}, M., {Meyer}, S.~S., {Page}, L., {Tucker}, G.~S., {Weiland}, J.~L.,
  {Wollack}, E., \& {Wright}, E.~L. 2003, \apjs, 148, 175

\bibitem[{{Springel}(2005)}]{2005MNRAS.364.1105S}
{Springel}, V. 2005, \mnras, 364, 1105

\bibitem[{{Tonry} {et~al.}(2000){Tonry}, {Blakeslee}, {Ajhar}, \&
  {Dressler}}]{2000ApJ...530..625T}
{Tonry}, J.~L., {Blakeslee}, J.~P., {Ajhar}, E.~A., \& {Dressler}, A. 2000,
  \apj, 530, 625

\bibitem[{{Tonry} \& {Davis}(1981)}]{1981ApJ...246..680T}
{Tonry}, J.~L. \& {Davis}, M. 1981, \apj, 246, 680

\bibitem[{{Tonry} {et~al.}(2001){Tonry}, {Dressler}, {Blakeslee}, {Ajhar},
  {Fletcher}, {Luppino}, {Metzger}, \& {Moore}}]{2001ApJ...546..681T}
{Tonry}, J.~L., {Dressler}, A., {Blakeslee}, J.~P., {Ajhar}, E.~A., {Fletcher},
  A.~B., {Luppino}, G.~A., {Metzger}, M.~R., \& {Moore}, C.~B. 2001, \apj, 546,
  681

\bibitem[{{Tully}(1988{\natexlab{a}})}]{1988lsmu.book..169T}
{Tully}, R.~B. 1988{\natexlab{a}}, {Distances to galaxies in the field}
  (Large-Scale Motions in the Universe: A Vatican study Week), 169--177

\bibitem[{{Tully}(1988{\natexlab{b}})}]{1988ngc..book.....T}
---. 1988{\natexlab{b}}, {Nearby galaxies catalog} (Cambridge and New York,
  Cambridge University Press, 1988, 221 p.)

\bibitem[{{Tully} \& {Fisher}(1977)}]{1977A&A....54..661T}
{Tully}, R.~B. \& {Fisher}, J.~R. 1977, \aap, 54, 661

\bibitem[{{Tully} \& {Fisher}(1987)}]{1987nga..book.....T}
---. 1987, {Nearby galaxies Atlas} (Cambridge: University Press, 1987)

\bibitem[{{Tully} \& {Fouqu{\'e}}(1985)}]{1985ApJS...58...67T}
{Tully}, R.~B. \& {Fouqu{\'e}}, P. 1985, \apjs, 58, 67

\bibitem[{{Tully} \& {Pierce}(2000)}]{2000ApJ...533..744T}
{Tully}, R.~B. \& {Pierce}, M.~J. 2000, \apj, 533, 744

\bibitem[{{Tully} {et~al.}(2006){Tully}, {Rizzi}, {Dolphin}, {Karachentsev},
  {Karachentseva}, {Makarov}, {Makarova}, {Sakai}, \&
  {Shaya}}]{2006AJ....132..729T}
{Tully}, R.~B., {Rizzi}, L., {Dolphin}, A.~E., {Karachentsev}, I.~D.,
  {Karachentseva}, V.~E., {Makarov}, D.~I., {Makarova}, L., {Sakai}, S., \&
  {Shaya}, E.~J. 2006, \aj, 132, 729

\bibitem[{{Tully} \& {Shaya}(1984)}]{1984ApJ...281...31T}
{Tully}, R.~B. \& {Shaya}, E.~J. 1984, \apj, 281, 31

\bibitem[{{Tully} {et~al.}(1992){Tully}, {Shaya}, \&
  {Pierce}}]{1992ApJS...80..479T}
{Tully}, R.~B., {Shaya}, E.~J., \& {Pierce}, M.~J. 1992, \apjs, 80, 479

\bibitem[{{van de Weygaert} \& {Schaap}(2007)}]{2007arXiv0708.1441V}
{van de Weygaert}, R. \& {Schaap}, W. 2007, ArXiv e-prints, 708

\bibitem[{{Yahil} {et~al.}(1977){Yahil}, {Tammann}, \&
  {Sandage}}]{1977ApJ...217..903Y}
{Yahil}, A., {Tammann}, G.~A., \& {Sandage}, A. 1977, \apj, 217, 903

\end{thebibliography}
